\pdfoutput=1
\documentclass[12pt]{article}\usepackage{qusann2}
\input{MyQcircuit.sty}
\begin{document}

\title{Code Generator\\
for Quantum Simulated Annealing}

\author{Robert R. Tucci\\
        P.O. Box 226\\
        Bedford,  MA   01730\\
        tucci@ar-tiste.com}

\date{ \today}

\maketitle

\vskip2cm
\section*{Abstract}

This paper introduces
QuSAnn v1.2 and Multiplexor Expander v1.2,
two Java applications
available for free. (Source code
included in the distribution.)
QuSAnn is a ``code generator" for
quantum simulated annealing:
after the user inputs some parameters,
it outputs
a quantum circuit
for  performing simulated annealing
on a quantum computer. The
quantum circuit
implements the algorithm
of Wocjan et al. (arXiv:0804.4259),
which improves on the original algorithm of
Somma et al. (arXiv:0712.1008).
The quantum circuit
generated by QuSAnn
includes some quantum multiplexors.
The application Multiplexor Expander
allows the user to replace
each of those multiplexors by
a sequence of more elementary
gates such as multiply
controlled NOTs and qubit rotations.

\section{Introduction}

For an explanation of the mathematical
notation
used in this paper, see some of my previous papers;
for instance, Ref.\cite{notation} Section 2.

We say a unitary operator
acting an array of qubits has been
compiled if
it has been expressed
as a SEO (Sequence of Elementary
Operations, like CNOTs and single-qubit operations).
SEO's are often represented as quantum circuits.

There exist software, ``general quantum compilers"
(like Qubiter\cite{TucQubiter}),
for compiling arbitrary unitary
operators (operators that have no a priori
known structure).
There also exists
software,``special purpose
quantum compilers"(like
each of the 7
applications
in QuanSuite\cite{quantree,quanfou,quanfruit}),
for
compiling unitary operators
that have a very definite, special
structure which is known a priori.

This paper introduces
QuSAnn v1.2 and Multiplexor Expander v1.2,
two Java applications
available for free. (Source code
included in the distribution.)
QuSAnn is a ``code generator"
for quantum simulated annealing:
after the user inputs some parameters,
it outputs
a quantum circuit
for performing simulated annealing
on a quantum computer.
QuSAnn is not really a quantum
compiler (neither general nor special)
because, although it
generates a quantum circuit like
the quantum compilers do,
it doesn't
start with an explicitly
stated unitary matrix
as input.
Multiplexor Expander is not
a quantum compiler either for the same reason.
Multiplexor Expander can be more aptly described
as a ``code translator":
it takes a SEO and replaces
it by a different but equivalent SEO.

In Ref.\cite{somma}, Somma et al.
proposed an algorithm for
quantum simulated annealing that
requires, for any $\epsilon>0$,
 order $1/\sqrt{\delta}$
elementary operations
to find, with probability greater than $1-\epsilon$,
the minimum of a
function.
Here $\delta$ is the distance
between the two largest
eigenvalue magnitudes
of the transition probability
matrix for the Metropolis Markov
chain. The algorithm of Somma et al.
outperforms the classical
simulated annealing algorithm,
which requires order $1/\delta$
elementary operations to do the same thing.

Subsequently, Wocjan et al.
in Ref.\cite{wocjan1}
improved on the algorithm
of Somma et al.
(See also
Refs.\cite{wocjan2,wocjanSparse},
where Wocjan et al. discuss related issues).

Both the Somma et al. and the Wocjan et al.
algorithms use Szegedy
quantum walk operators\cite{szegedy}
and phase estimation.
But the Wocjan et al. algorithm uses
a Grover fixed point search\cite{grover}
instead of the quantum Zeno effect.

The
quantum circuit generated by QuSAnn
implements the algorithm
of Wocjan et al. given in Ref.\cite{wocjan1}.
The circuit
includes some quantum multiplexors.
(See Ref.\cite{notation}
for a review of quantum multiplexors.)
The application Multiplexor Expander
allows the user to replace
each of those multiplexors by
a sequence of more elementary
gates such as multiply
controlled NOTs and qubit rotations.
Multiplexor Expander
gives the user the option of expanding the
multiplexors in two
different ways: either as
an Exact SEO
(see Refs. \cite{TucQubiter,notation}),
or
an Oracular Approximation
(see Ref. \cite{TucOracularApp}).

A nice feature of the
source code for QuSAnn,
Multiplexor Expander and
all the applets in QuanSuite,
is that they all share a common Java
class library (named QLib).

\section{QuSAnn}

The QuSAnn applet makes the following
3 assumptions:

\begin{enumerate}
\item
For its {\it annealing schedule} (see Appendix
\ref{app-class-sim-ann} for definition)
$\beta_0,\beta_1, \beta_2,\ldots \beta_{t_f}$,
it assumes $\beta_0=0$ and
$\beta_{j+1}-\beta_j=\Delta\beta>0$ for
all $j=0,1,\ldots,t_f-1$.

\item
For its {\it energy function} (see Appendix
\ref{app-class-sim-ann} for definition),
it assumes $E(x)= (x-\frac{\ns}{2})^2$,
where the state space (i.e., the
set of $x$ values) of the minimization
problem is $\{0,1,2,\dots,\ns-1\}$.

\item
For its {\it neighborhood function} (see Appendix
\ref{app-class-sim-ann} for definition),
it assumes
$neig(x,y) = 1$ if $|x-y|\leq 1$
and $neig(x,y) = 0$ otherwise.
\end{enumerate}

These 3 assumptions
were made in order
to make
the applet simple. They
can be
easily changed (i.e, one
can use a more complicated
annealing schedule, energy function
and neighborhood function)
by making trivial alterations to
the source code of QuSAnn.

\subsection{The Control Panel}

\begin{figure}[h]
    \begin{center}
    \includegraphics[scale=.70]{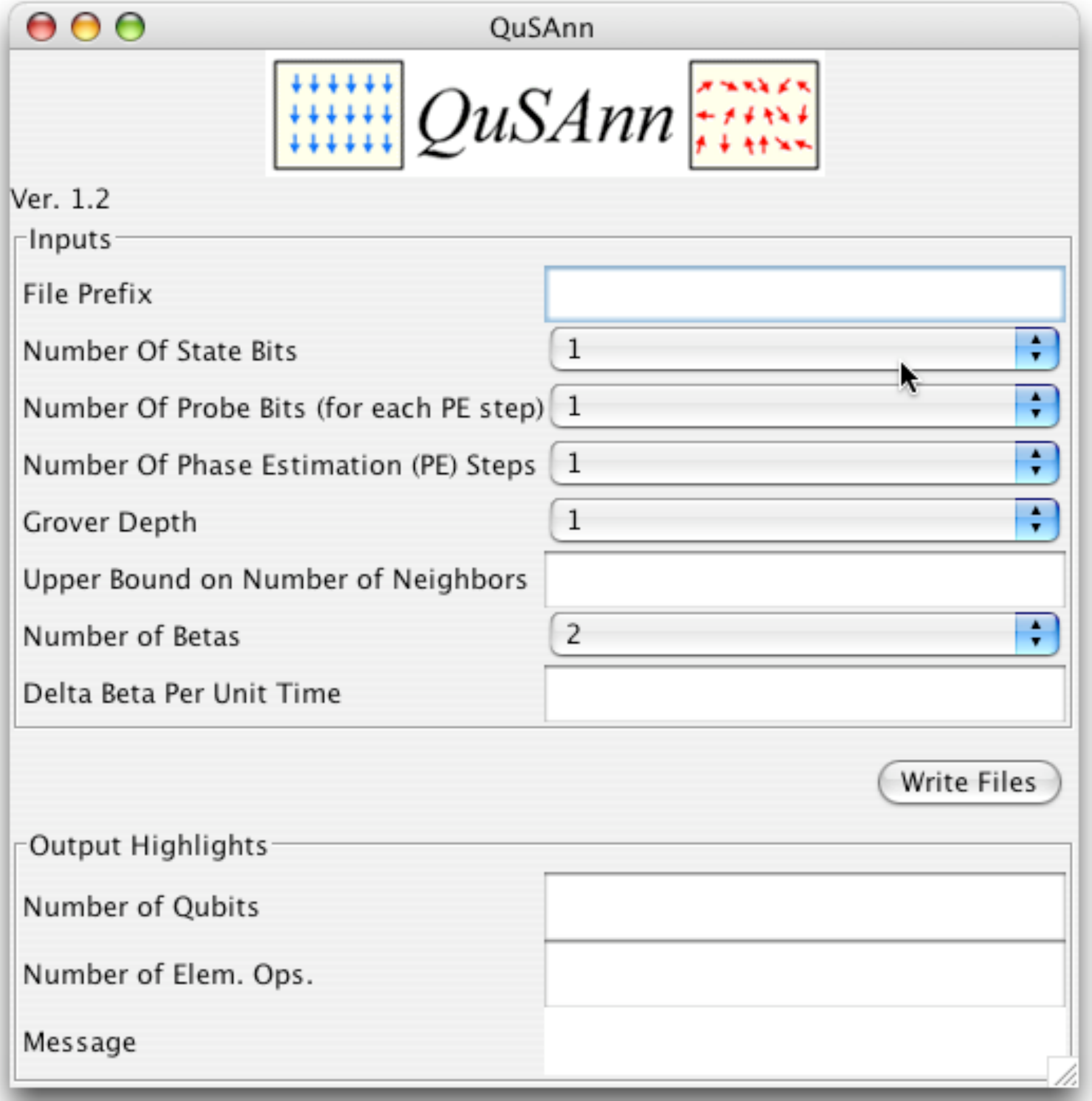}
    \caption{Control Panel of QuSAnn}
    \label{fig-qusann-main}
    \end{center}
\end{figure}

Fig.\ref{fig-qusann-main} shows the
{\bf Control Panel} for QuSAnn. This is the
main and only window of the application. This
window is
open if and only if the application is running.
\newpage

The Control Panel
allows you to enter the following inputs:
\begin{description}

\item[File Prefix:]Prefix to the
3 output files
that are written when you press the
{\bf Write Files} button. For example,
if you insert {\tt test} in this text field,
the following 3 files will be written:
\begin{itemize}
\item
{\tt test\_qsann\_log.txt}
\item
{\tt test\_qsann\_eng.txt}
\item
{\tt test\_qsann\_pic.txt}
\end{itemize}
Some examples of these output files
are given in Section \ref{sec-sann-out-files}
below.

\item[Number of State Bits:]
The parameter
$\nb=1,2,3\ldots$ defined in Appendix
\ref{app-class-sim-ann}. The
state space of the
minimization problem is
$\{0,1,2,\ldots, 2^\nb-1\}$.
The transition probability
matrix $M$ of the Metropolis Markov
chain is a $2^\nb$ dimensional matrix,
and the Szegedy quantum walk operator
$W(M)$ is a $2^{2\nb}$
dimensional matrix.

\item[Number of Probe Bits (for each PE step):]
The parameter $a=1,2,3,\ldots$ defined in Appendix
\ref{app-wa-algo}. See Fig.\ref{fig-v-beta}.

\item[Number of Phase Estimation (PE) Steps:]
The parameter $c=1,2,3,\ldots$ defined in Appendix
\ref{app-wa-algo}. See Fig.\ref{fig-v-beta}.

\item[Grover Depth:]
The parameter $d_f=1,2,3,\ldots$ defined in Appendix
\ref{app-wa-algo}. $d_f$ is the final level of recursion
to which one wishes to carry out
the fixed point Grover search.

\item[Upper Bound on Number of Neighbors:]
The parameter
$upBdNeig\in \RR^{>0}$ defined in Appendix
\ref{app-class-sim-ann}.

\item[Number of Betas:]
The parameter $t_f+1=2,3,\ldots$ defined above
to be the number of betas in
the annealing schedule.

\item[Delta Beta Per Unit Time:]
The parameter $\Delta\beta\in \RR^{>0}$
defined above to be the difference
between adjacent betas of
the annealing schedule.
\end{description}

The Control Panel displays the
 following
outputs:
\begin{description}
\item[Number of Qubits:] The total
number of qubits used by the circuit,
equal to $2\nb+ac$ in the notation of
Appendix \ref{app-wa-algo}.

\item[Number of Elementary Operations:]
The number of elementary operations
in the output quantum circuit.
If there are no LOOPs, this is
the number of lines in the English File
(see Sec. \ref{sec-eng-file}), which
equals the number of lines in the
Picture File (see Sec. \ref{sec-pic-file}).
For a LOOP (which is not nested
inside a larger LOOP), the
``{\tt LOOP k REPS:$N$}" and
``{\tt NEXT k}" lines are not counted,
whereas the lines between
``{\tt LOOP k REPS:$N$}" and
``{\tt NEXT k}"
are counted $N$ times.
Multiplexors expressed as a
single line are counted as a
single elementary operation
(unless, of course, they are inside a LOOP,
in which case they are used repeatedly).

\item[Message:]
A message appears in this text field
if you press
{\bf Write Files} with a bad input.
The message tries to explain
the mistake in the input.

\end{description}

\subsection{Output Files}
\label{sec-sann-out-files}

\begin{figure}[h]
    \begin{center}
    \includegraphics[scale=.70]{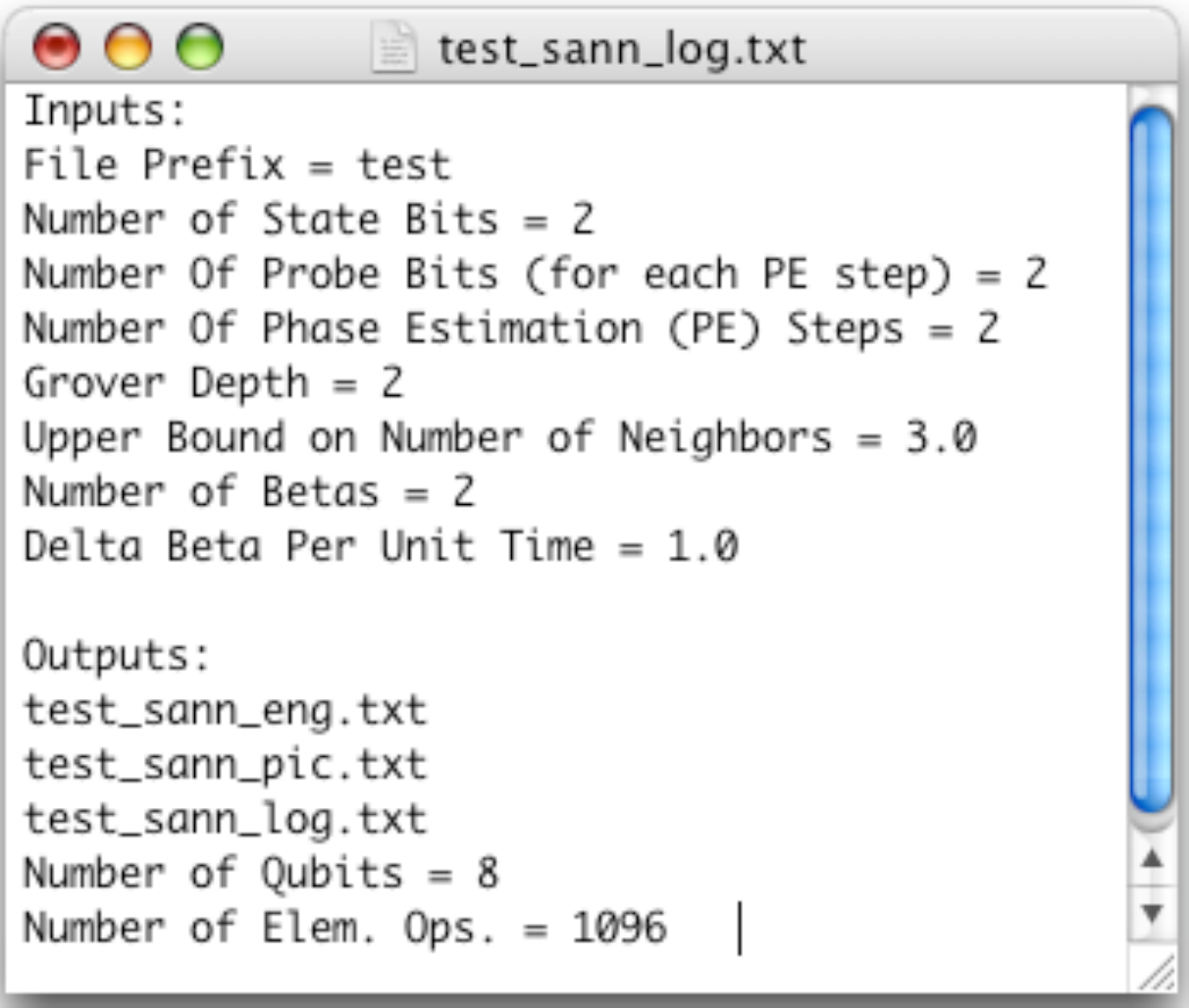}
    \caption{Log File generated by QuSAnn}
    \label{fig-sann-log}
    \end{center}
\end{figure}

\begin{figure}[h]
    \begin{center}
    \includegraphics[scale=.70]{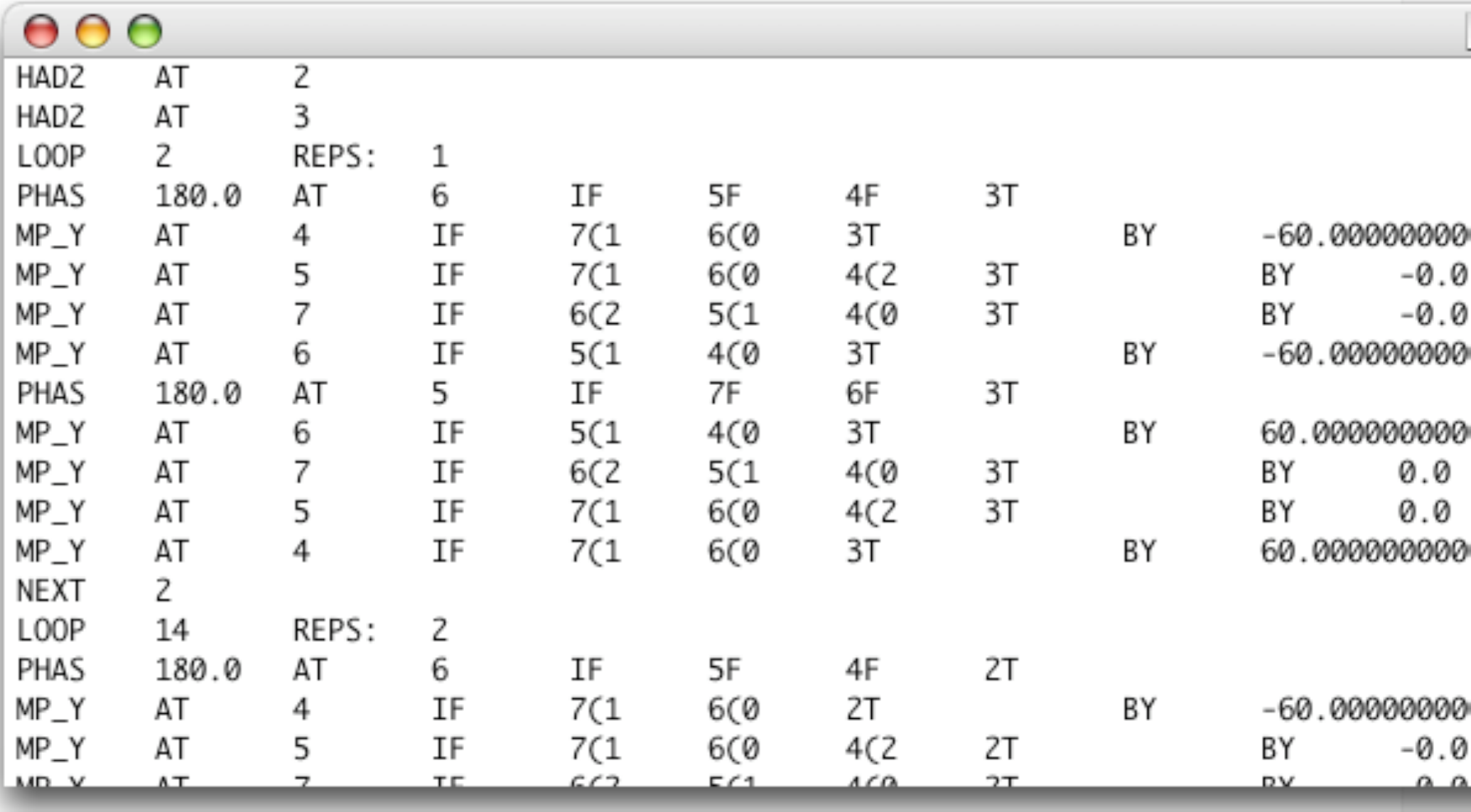}
    \caption{English File generated by
    QuSann in the same run as the
     Log File of Fig.\ref{fig-sann-log}.
     Bottom of file is not visible.
     Right hand side of file is not visible.
    }
    \label{fig-sann-eng}
    \end{center}
\end{figure}

\begin{figure}[h]
    \begin{center}
    \includegraphics[scale=.70]{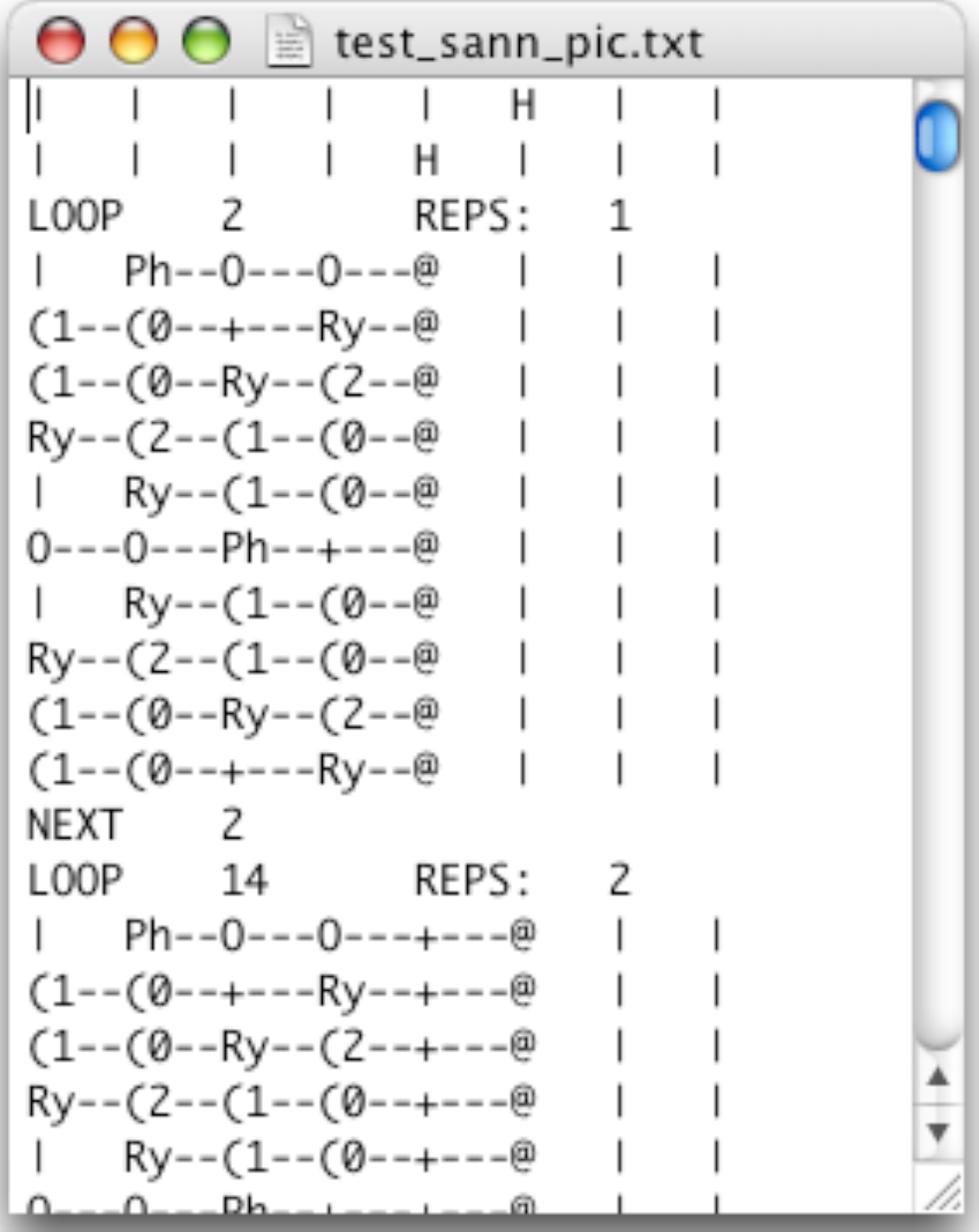}
    \caption{Picture File generated by
    QuSAnn in the same run as the
     Log File of Fig.\ref{fig-sann-log}.
     Bottom of file is not visible.
    }
    \label{fig-sann-pic}
    \end{center}
\end{figure}

Figs. \ref{fig-sann-log},
\ref{fig-sann-eng} and
\ref{fig-sann-pic},
were all generated
in a single run
of QuSAnn (by pressing
the {\bf Write Files} button just once).
They are examples of what we call the {\bf
Log File, English File, and Picture File}, respectively, of QuSAnn.
Next we explain
the contents of each of these output files.

\newpage

\subsubsection{Log File}
Fig.\ref{fig-sann-log}
is an example a Log File.
The Log File
records all the information
found in the Control Panel.

\subsubsection{English File}\label{sec-eng-file}

Fig.\ref{fig-sann-eng}
is an example of an English File.
The English File
completely specifies the output SEO.
It does so ``in English", thus its name.
Each line represents one elementary operation,
and time increases as we move downwards.

In general, an English File obeys
the following rules:

\begin{itemize}
\item Time grows as we move down the file.

\item Each row
corresponds to one elementary operation.
Each row starts with 4 letters that indicate
the type of elementary operation.

\item For a one-bit operation
acting on a ``target bit" $\alpha$,
the target bit $\alpha$ is given after the word {\tt AT}.

\item If the one-bit operation is controlled, then
the controls are indicated after the word {\tt IF}.
{\tt T} and {\tt F} stand for
true and false, respectively.
{\tt $\alpha$T} stands for
a control $n(\alpha)$ at bit $\alpha$.
{\tt $\alpha$F} stands for
a control $\nbar(\alpha)$ at bit $\alpha$.

\item ``{\tt LOOP k REPS:$N$}" and ``{\tt NEXT k}"
mark the beginning and end of $N$
iterations. {\tt k} labels the loop. {\tt k} also
equals the line-count number (first line is 0)
of the line
``{\tt LOOP k REPS:$N$}" in the English file.

\item {\tt SWAP
$\alpha$ $\beta$}
stands for the swap(exchange) operator
$E(\alpha, \beta)$
that swaps
bits $\alpha$ and $\beta$.

\item {\tt PHAS }$\theta^{degs}$ stands for
a phase factor $e^{i \theta^{degs} \frac{\pi}{180}}$.

\item
{\tt P0PH }$\theta^{degs}$ stands for
the one-bit gate
 $e^{i P_0 \theta^{degs} \frac{\pi}{180}}$.
{\tt P1PH }$\theta^{degs}$ stands for
  the one-bit gate
 $e^{i P_1 \theta^{degs} \frac{\pi}{180}}$.
 Target bit follows the word {\tt AT}.

\item {\tt SIGX}, {\tt SIGY},
{\tt SIGZ}, {\tt HAD2}
stand for
the Pauli matrices $\sigx, \sigy, \sigz$
and the one-bit Hadamard matrix $H$,
respectively.
Target bit follows the word {\tt AT}.

\item {\tt ROTX}, {\tt ROTY},
{\tt ROTZ}, {\tt ROTN}
stand for one-bit
rotations
with rotation axes in the
directions: $x$, $y$, $z$, and
an arbitrary direction $n$, respectively.
Rotation angles (in degrees) follow
the words {\tt ROTX}, {\tt ROTY},
{\tt ROTZ}, {\tt ROTN}.
Target bit follows the word {\tt AT}.

\item
{\tt MP\_Y} stands for a multiplexor
which performs a one-bit rotation of
a target bit about the
$y$ axis. Target bit follows the word {\tt AT}.
Rotation angles (in degrees) follow
the word {\tt BY}. Multiplexor controls
are specified by $\alpha(k$, where
integer $\alpha$ is the bit position
and integer $k$ is the control's name.

\end{itemize}

Here is a list of examples
showing how to translate the mathematical
notation used in Ref.\cite{notation}
into the English File language:\footnote{
This same table appeared before
in Ref.\cite{quantree}
except that we have
added a new row of boxes at the end of the table
for multiplexors expressed as
a single line in the English
and Picture files.}
\begin{center}
\begin{tabular}{|l|l|}
\hline
Mathematical language & English File language\\
\hline
\hline
Loop named 5 with 2 repetitions &
{\tt  LOOP 5 REPS: 2}\\
\hline
Next iteration of loop named 5&
{\tt  NEXT 5}\\
\hline
$E(1,0)^{\nbar(3)n(2)}$ &
{\tt SWAP  1  0  IF  3F  2T}\\
\hline
$e^{i 42.7 \frac{\pi}{180} \nbar(3)n(2)}$ &
{\tt  PHAS 42.7 IF  3F  2T}\\
\hline
$e^{i 42.7 \frac{\pi}{180} \nbar(3)n(2)}$ &
{\tt  P0PH 42.7 AT  3 IF 2T}\\
\hline
$e^{i 42.7 \frac{\pi}{180} n(3)n(2)}$ &
{\tt  P1PH 42.7 AT  3 IF 2T}\\
\hline
$\sigx(1)^{\nbar(3)n(2)}$ &
{\tt  SIGX  AT  1  IF  3F  2T}\\
\hline
$\sigy(1)^{\nbar(3)n(2)}$ &
{\tt  SIGY  AT  1  IF  3F  2T}\\
\hline
$\sigz(1)^{\nbar(3)n(2)}$ &
{\tt  SIGZ  AT  1  IF  3F  2T}\\
\hline
$H(1)^{\nbar(3)n(2)}$ &
{\tt  HAD2  AT  1  IF  3F  2T}\\
\hline
$(e^{\frac{i}{2} \frac{\pi}{180} 23.7 \sigx(1)})^{\nbar(3)n(2)}$ &
{\tt  ROTX  23.7  AT  1  IF  3F  2T}\\
\hline
$(e^{\frac{i}{2}  \frac{\pi}{180} 23.7 \sigy(1)})^{\nbar(3)n(2)}$ &
{\tt  ROTY  23.7  AT  1  IF  3F  2T} \\
\hline
$(e^{\frac{i}{2}  \frac{\pi}{180} 23.7 \sigz(1)})^{\nbar(3)n(2)}$ &
{\tt  ROTZ  23.7  AT  1  IF  3F  2T}\\
\hline
$(e^{
\frac{i}{2}  \frac{\pi}{180}
[30\sigx(1)+ 40\sigy(1) + 11 \sigz(1)]
 })^{\nbar(3)n(2)}$ &
{\tt  ROTN  30.0 40.0 11.0  AT  1  IF  3F  2T}\\
\hline
$[e^{i\sum_{b1,b0}\theta_{b_1b_0}\sigy(3)P_{b_1b_0}(2,1)}]^{n(0)}$
&
{\tt MP\_Y  AT  3 IF 2(1 1(0 0T BY 30.0 10.5 11.0 83.1}
\\
where $\left\{\begin{array}{l}
\theta_{00}=30.0(\frac{\pi}{180})
\\
\theta_{01}=10.5(\frac{\pi}{180})
\\
\theta_{10}=11.0(\frac{\pi}{180})
\\
\theta_{11}=83.1(\frac{\pi}{180})
\end{array}\right.$
&\;
\\
\hline
\end{tabular}
\end{center}

\subsubsection{ASCII Picture File}\label{sec-pic-file}

Fig.\ref{fig-sann-pic}
is an example of a Picture File.
The Picture File
partially specifies the output SEO.
It gives an ASCII picture of
the quantum circuit.
Each line represents one elementary operation,
and time increases as we move downwards.
There is a one-to-one onto correspondence
between the rows of the English
and Picture Files.

In general, a Picture File obeys
the following rules:

\begin{itemize}
\item Time grows as we move down the file.

\item Each row
corresponds to one elementary operation.
Columns $1, 5, 9, 13, \ldots$ represent
qubits (or, qubit positions). We define the
rightmost qubit  as 0. The qubit
immediately to
the left of the rightmost qubit
is 1, etc.
For a one-bit operator
acting on a ``target bit" $\alpha$,
one places a symbol
of the operator at bit position
$\alpha$.

\item {\tt |} represents a ``qubit wordline"
connecting the same qubit at
two consecutive times.

\item {\tt -}represents a wire connecting different
qubits at the same time.

\item{\tt +} represents both {\tt |} and {\tt -}.

\item If the one-bit operation is controlled, then
the controls are indicated
as follows.
{\tt @} at bit position $\alpha$ stands for
a control $n(\alpha)$.
{\tt 0} at bit position $\alpha$ stands for
a control $\nbar(\alpha)$.

\item ``{\tt LOOP k REPS:$N$}" and ``{\tt NEXT k}"
mark the beginning and end of $N$
iterations. {\tt k} labels the loop. {\tt k} also
equals the line-count number (first line is 0)
of the line
``{\tt LOOP k REPS:$N$}" in the Picture File.

\item The swap(exchange) operator
$E(\alpha, \beta)$
is represented by putting
arrow heads {\tt <} and {\tt >} at
bit positions $\alpha$ and $\beta$.

\item A
phase factor $e^{i\theta}$ for
$\theta\in \RR$ is represented by
placing {\tt Ph} at any bit position
which does not already hold a control.

\item The one-bit gate
$e^{i P_0(\alpha)\theta}$ for $\theta\in \RR$
is represented by putting {\tt OP}
at bit position $\alpha$.

\item The one-bit gate
$e^{i P_1(\alpha)\theta}$ for $\theta\in \RR$
is represented by putting {\tt @P}
at bit position $\alpha$.

\item One-bit operations
 $\sigx(\alpha)$,
 $\sigy(\alpha)$,
 $\sigz(\alpha)$
and $H(\alpha)$
are represented by placing the letters
{\tt X,Y,Z, H}, respectively,
at bit position $\alpha$.

\item
One-bit rotations
acting on bit $\alpha$,
in the
$x,y,z,n$ directions,
are represented by placing
{\tt Rx,Ry,Rz, R}, respectively,
at bit position $\alpha$.

\item
A multiplexor that rotates
a bit $\tau$ about the $y$ axis
is represented by placing
{\tt Ry}
at bit position $\tau$.
A multiplexor control at bit position $\alpha$
and
named by the integer $k$
is represented by placing
$(k$ at bit position $\alpha$.
\end{itemize}

Here is a list of examples
showing how to translate the mathematical
notation used in Ref.\cite{notation}
into the Picture File language:\footnote{
This same table appeared before
in Ref.\cite{quantree}
except that we have
added a new row of boxes at the end of the table
for multiplexors expressed as
a single line in the English
and Picture files.}

\begin{tabular}{|l|l|}
\hline
Mathematical language & Picture File language\\
\hline
\hline
Loop named 5 with 2 repetitions &
{\tt  LOOP 5 REPS:2}\\
\hline
Next iteration of loop named 5&
{\tt  NEXT 5}\\
\hline
$E(1,0)^{\nbar(3)n(2)}$& {\tt 0---@---<--->} \\
\hline
$e^{i 42.7 \frac{\pi}{180} \nbar(3)n(2)}$ &
{\tt 0---@---+--Ph}\\
\hline
$e^{i 42.7 \frac{\pi}{180} \nbar(3)n(2)}$ &
{\tt 0P--@\ \ \ |\ \ \ |}\\
\hline
$e^{i 42.7 \frac{\pi}{180} n(3)n(2)}$ &
{\tt @P--@\ \ \ |\ \ \ |}\\
\hline

 $\sigx(1)^{\nbar(3)n(2)}$& {\tt 0---@---X\ \ \ |} \\
\hline
 $\sigy(1)^{\nbar(3)n(2)}$& {\tt 0---@---Y\ \ \ |} \\
\hline
 $\sigz(1)^{\nbar(3)n(2)}$& {\tt 0---@---Z\ \ \ |} \\
\hline
$H(1)^{\nbar(3)n(2)}$& {\tt 0---@---H\ \ \ |} \\
\hline
$(e^{\frac{i}{2} \frac{\pi}{180} 23.7 \sigx(1)})^{\nbar(3)n(2)}$&
{\tt 0---@---Rx\ \ |} \\
\hline
$(e^{\frac{i}{2}  \frac{\pi}{180} 23.7 \sigy(1)})^{\nbar(3)n(2)}$&
{\tt 0---@---Ry\ \ |} \\
\hline
$(e^{\frac{i}{2}  \frac{\pi}{180} 23.7 \sigz(1)})^{\nbar(3)n(2)}$&
{\tt 0---@---Rz\ \ |} \\
\hline
$(e^{
\frac{i}{2}  \frac{\pi}{180}
[30\sigx(1)+ 40\sigy(1) + 11 \sigz(1)]
})^{\nbar(3)n(2)}$&
{\tt 0---@---R\ \ \ |} \\
\hline
$[e^{i\sum_{b1,b0}\theta_{b_1b_0}\sigy(3)P_{b_1b_0}(2,1)}]^{n(0)}$
&
{\tt |\ \ \ Ry--(1--(0--@}
\\
where $\left\{\begin{array}{l}
\theta_{00}=30.0(\frac{\pi}{180})
\\
\theta_{01}=10.5(\frac{\pi}{180})
\\
\theta_{10}=11.0(\frac{\pi}{180})
\\
\theta_{11}=83.1(\frac{\pi}{180})
\end{array}\right.$
&\;
\\
\hline
\end{tabular}

\section{Multiplexor Expander}
QuSAnn outputs a quantum
circuit which includes multiplexor operations.
Multiplexor Expander can
read the output files of QuSAnn
and
write new files
in which each multiplexor
is replaced by a sequence of more
elementary operations such as
multiply controlled NOTs and single
qubit rotations.
Multiplexor Expander
gives the user the option of expanding the
multiplexors in two
different ways: either as
an Exact SEO
(see Ref. \cite{TucQubiter,notation}),
or
an Oracular Approximation
(see Ref. \cite{TucOracularApp}).

\subsection{The Control Panel}

\begin{figure}[h]
    \begin{center}
    \includegraphics[scale=.70]{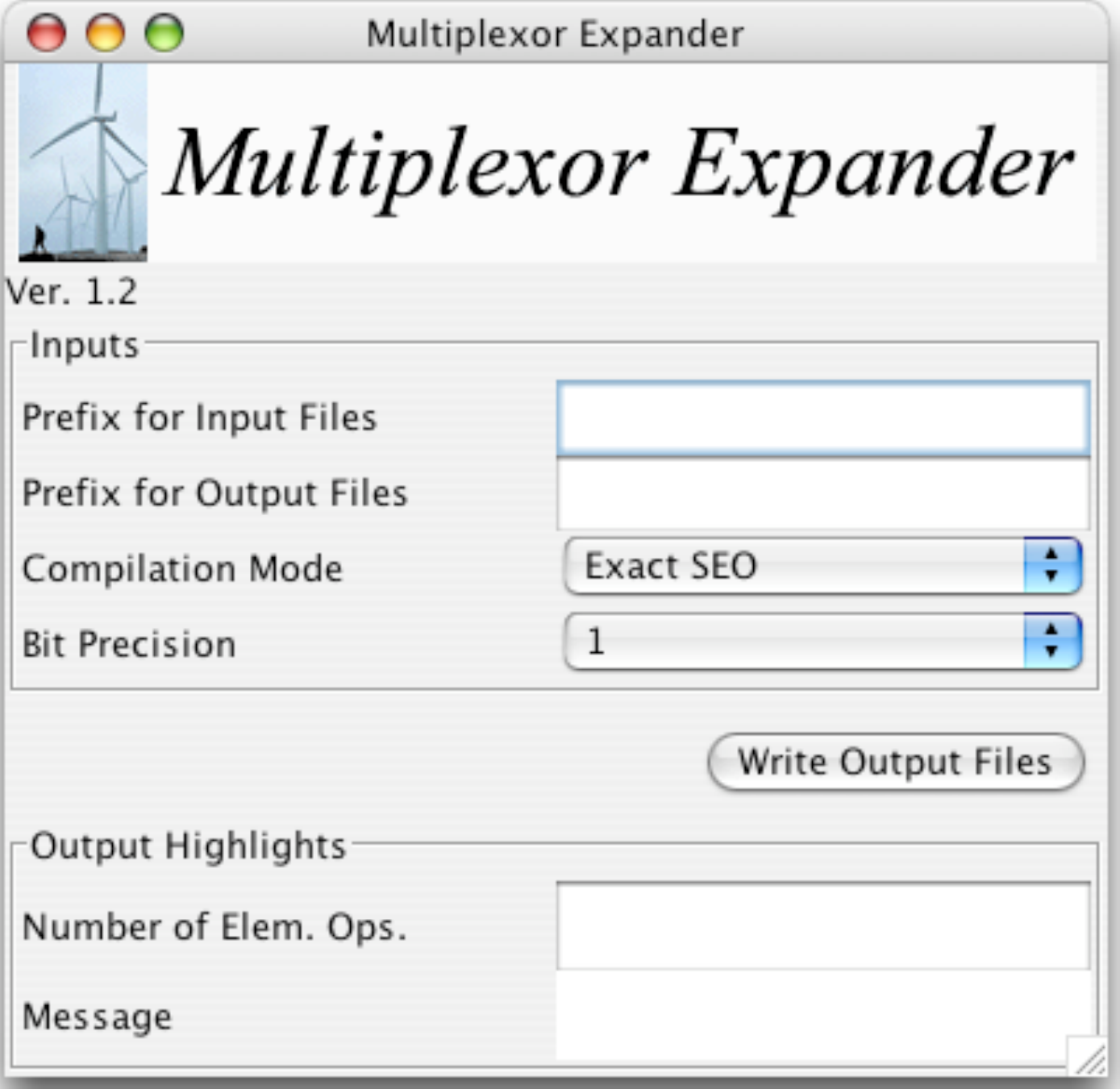}
    \caption{Control Panel of Multiplexor Expander}
    \label{fig-qexp-main}
    \end{center}
\end{figure}

Fig.\ref{fig-qexp-main} shows the
{\bf Control Panel} for Multiplexor Expander.
This is the
main and only window of the application. This
window is
open if and only if the application is running.

The Control Panel
allows you to enter the following inputs:
\begin{description}
\item[Prefix for Input Files:]
Prefix to the
2 input files that are read when you press the
{\bf Write Output Files} button.
For example,
if you insert {\tt testIn\_qexp} in this text field,
the following 2 files will be read:
\begin{itemize}
\item
{\tt testIn\_qexp\_eng.txt}
\item
{\tt testIn\_qexp\_pic.txt}
\end{itemize}
Some examples of these input files
are given in Section \ref{sec-qexp-in-files}
below.
These 2 files are usually
English File and Picture File
outputted by QuSAnn. If
they aren't, they must
be formatted in the same way
as the 2 files outputted by
QuSAnn or else Multiplexor Expander will fail.

\item[Prefix for Output Files:]
Prefix to the
3 output files
that are written when you press the
{\bf Write Output Files} button. For example,
if you insert {\tt testOut\_qexp} in this text field,
the following 3 files will be written:
\begin{itemize}
\item
{\tt testOut\_qexp\_log.txt}
\item
{\tt testOut\_qexp\_eng.txt}
\item
{\tt testOut\_qexp\_pic.txt}
\end{itemize}
Some examples of these output files
are given in Section \ref{sec-qexp-out-files}
below.

\item[Compilation Mode:] The compilation
mode, either as Exact SEO
(see Ref. \cite{TucQubiter,notation}),
or
Oracular Approximation
(see Ref. \cite{TucOracularApp}).

\item[Bit Precision:] The number of significant
fractional bits in the oracular approximation
(see Ref.\cite{TucOracularApp}). This parameter
is ignored if the compilation
mode is Exact SEO.

\end{description}

The Control Panel displays the
 following
outputs:
\begin{description}

\item[Number of Elementary Operations:]
Same as in QuSAnn Control Panel.

\item[Message:]
Same as in QuSAnn Control Panel.

\end{description}

\subsection{Input Files}
\label{sec-qexp-in-files}

\begin{figure}[h]
    \begin{center}
    \includegraphics[scale=.70]{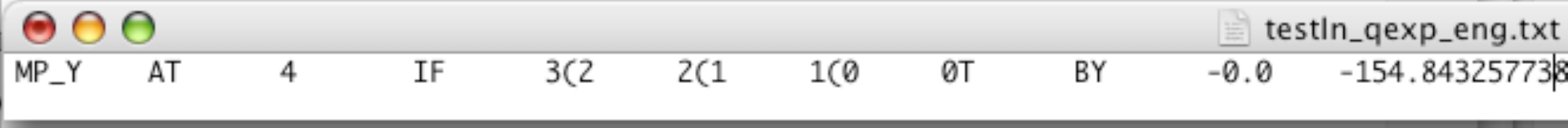}
    \caption{Input English File for Multiplexor Expander.
     Right hand side of file is not visible.}
    \label{fig-qexp-eng}
    \end{center}
\end{figure}

\begin{figure}[h]
    \begin{center}
    \includegraphics[scale=.70]{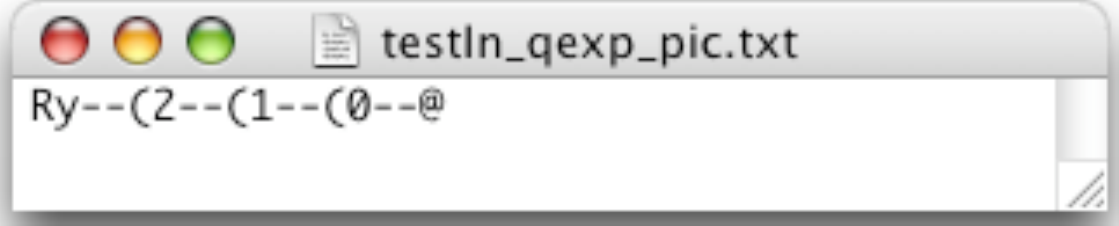}
    \caption{Input Picture File for Multiplexor Expander.}
    \label{fig-qexp-pic}
    \end{center}
\end{figure}

Figs.
\ref{fig-qexp-eng} and
\ref{fig-qexp-pic}
are examples of the 2 input files for
Multiplexor Expander,
what we call the {\bf Input English File}
and
{\bf Input Picture File}, respectively,
 of Multiplexor Expander.
These examples are not really
output files of QuSAnn,
but they are formatted
in the same way as English and
Picture files of QuSAnn.


\subsection{Output Files}
\label{sec-qexp-out-files}

\begin{figure}[h]
    \begin{center}
    \includegraphics[scale=.70]{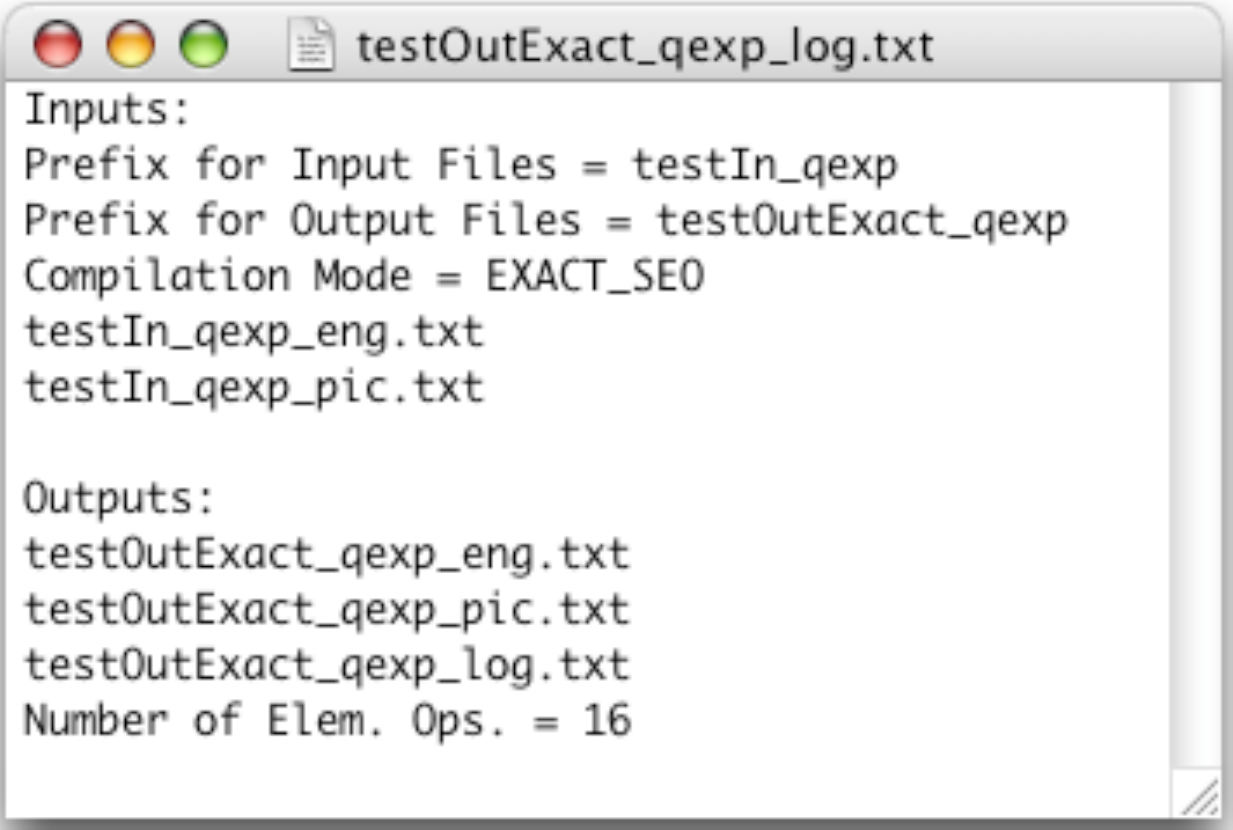}
    \caption{Output Log File generated by Multiplexor
    Expander with Exact SEO as
    compilation mode.}
    \label{fig-outExact-qexp-log}
    \end{center}
\end{figure}

\begin{figure}[h]
    \begin{center}
    \includegraphics[scale=.70]{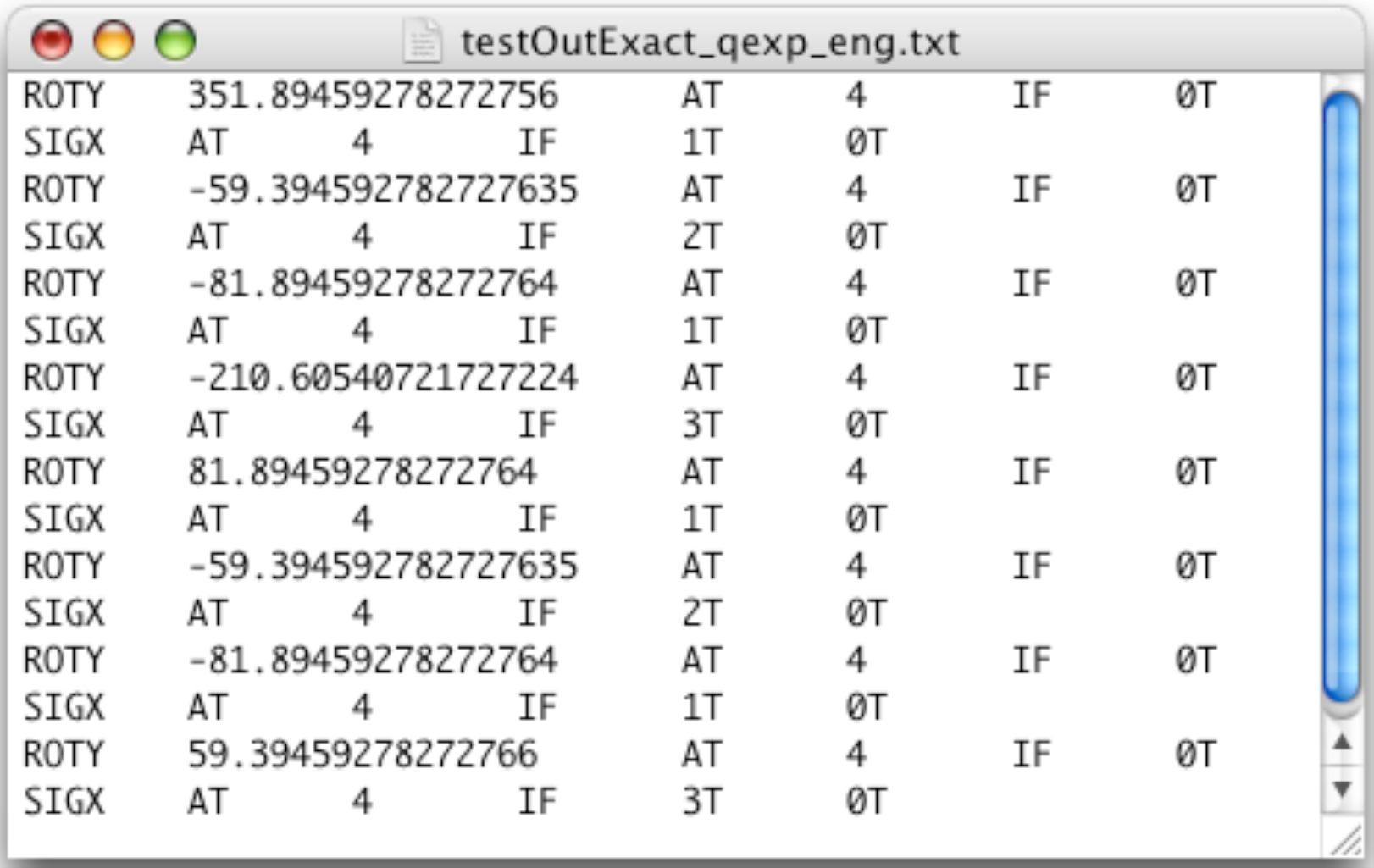}
    \caption{Output English File generated by
    Multiplexor Expander in the same run as the
     Log File of Fig.\ref{fig-outExact-qexp-log}}
    \label{fig-outExact-qexp-eng}
    \end{center}
\end{figure}

\begin{figure}[h!]
    \begin{center}
    \includegraphics[scale=.70]{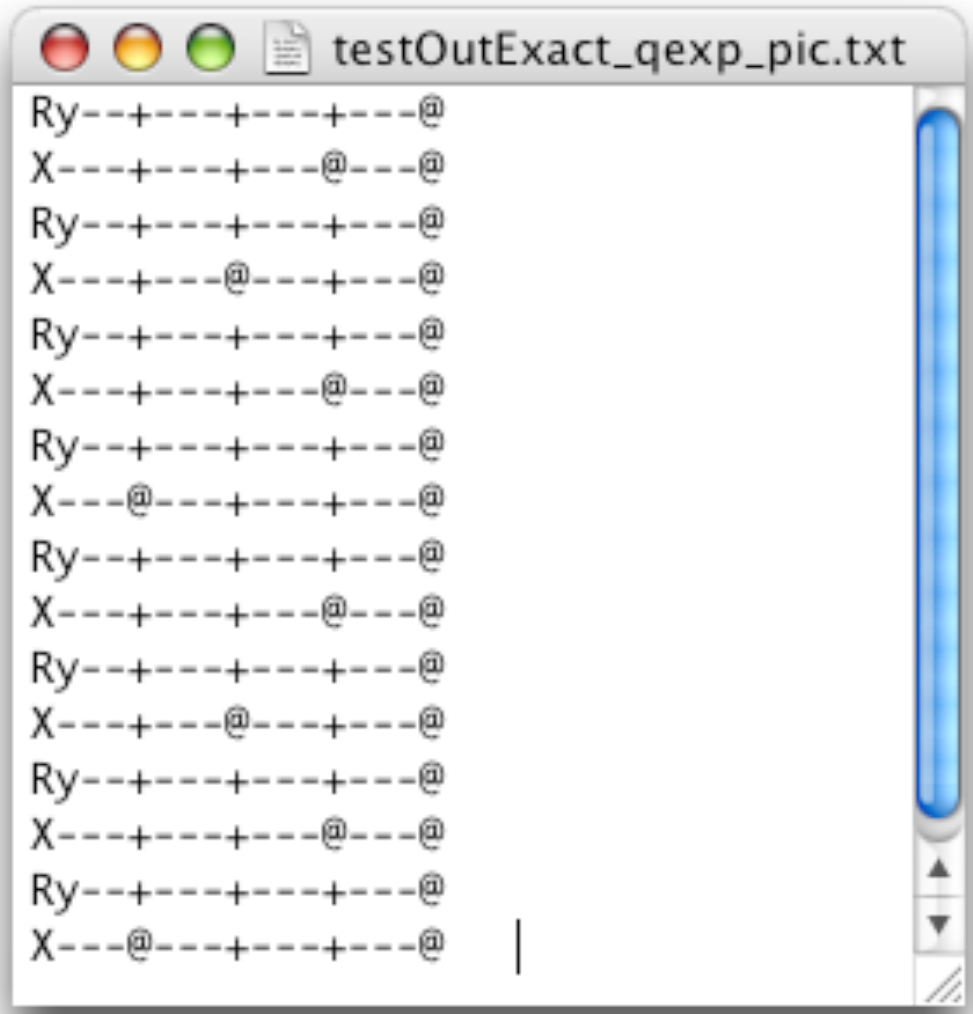}
    \caption{Output Picture File generated by
    Multiplexor Expander in the same run as the
     Log File of Fig.\ref{fig-outExact-qexp-log}}
    \label{fig-outExact-qexp-pic}
    \end{center}
\end{figure}

Figs. \ref{fig-outExact-qexp-log},
\ref{fig-outExact-qexp-eng} and
\ref{fig-outExact-qexp-pic}
were all generated
in a single run
of Multiplexor Expander (by pressing
the {\bf Write Output Files} button just once),
with
\begin{itemize}
\item
the input files of Figs. \ref{fig-qexp-eng}
and \ref{fig-qexp-pic}, and
\item
the compilation mode set to {\bf Exact SEO} .
\end{itemize}
Figs. \ref{fig-outExact-qexp-log},
\ref{fig-outExact-qexp-eng} and
\ref{fig-outExact-qexp-pic}
 are examples of what we call the {\bf
Output Log File, Output English File, and Output Picture File}, respectively,
of Multiplexor Expander.
The notation of these files is the same as that
for the Log, English and Picture files
for QuSAnn (see
Section \ref{sec-sann-out-files}).

\begin{figure}[h!]
    \begin{center}
    \includegraphics[scale=.70]{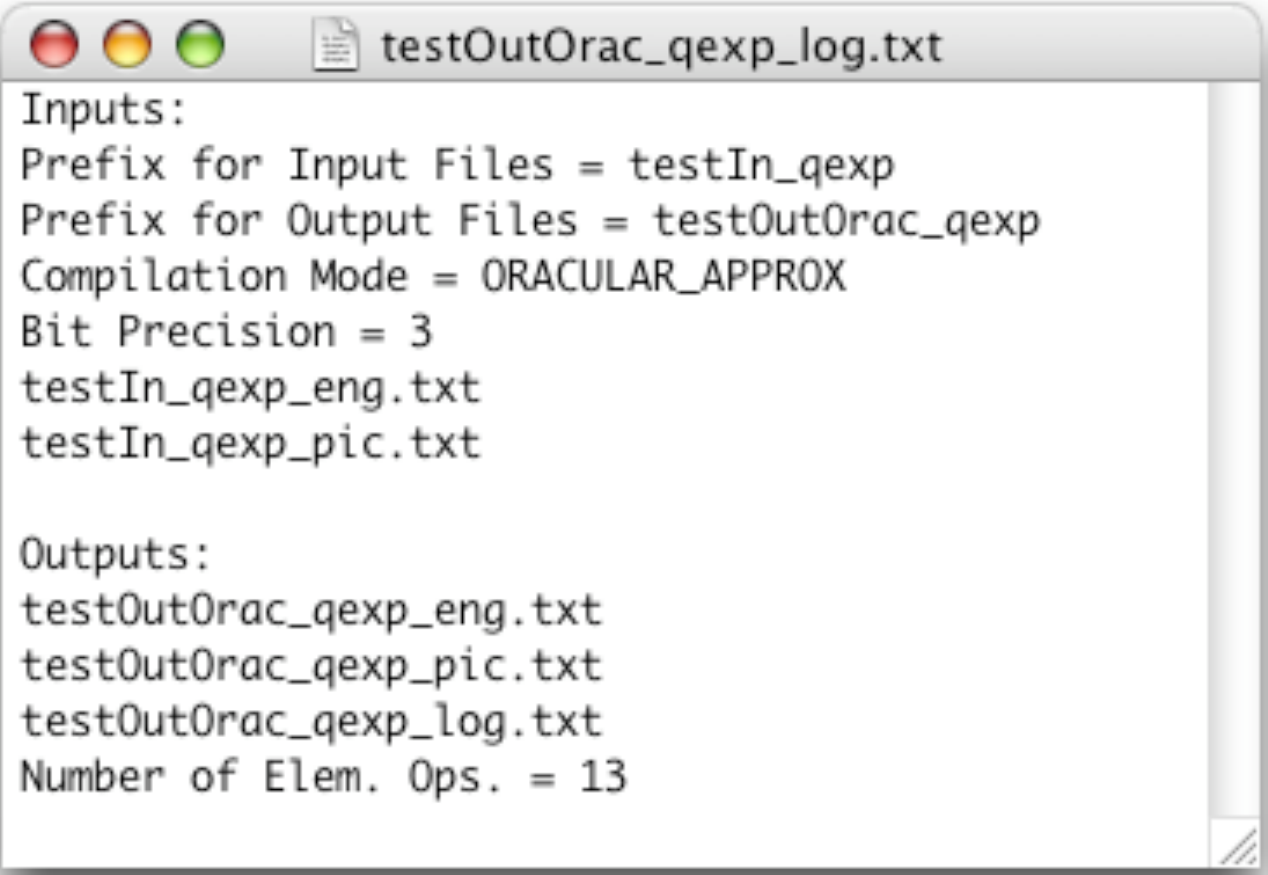}
    \caption{Output Log File generated by Multiplexor
    Expander with Oracular Approximation as
    compilation mode.}
    \label{fig-outOrac-qexp-log}
    \end{center}
\end{figure}

\begin{figure}[h!]
    \begin{center}
    \includegraphics[scale=.70]{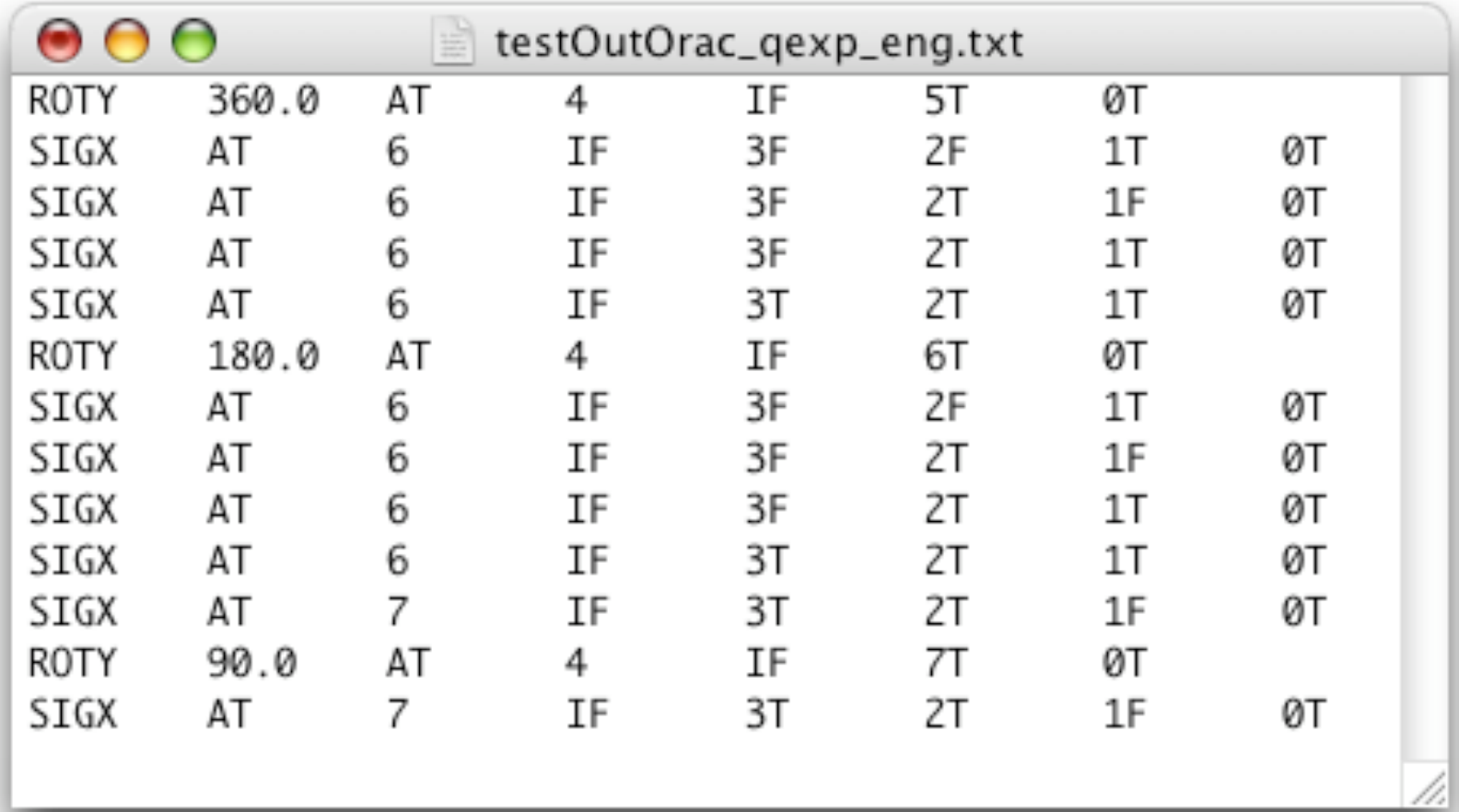}
    \caption{Output English File generated by
    Multiplexor Expander in the same run as the
     Log File of Fig.\ref{fig-outOrac-qexp-log}}
    \label{fig-outOrac-qexp-eng}
    \end{center}
\end{figure}

\begin{figure}[h!]
    \begin{center}
    \includegraphics[scale=.70]{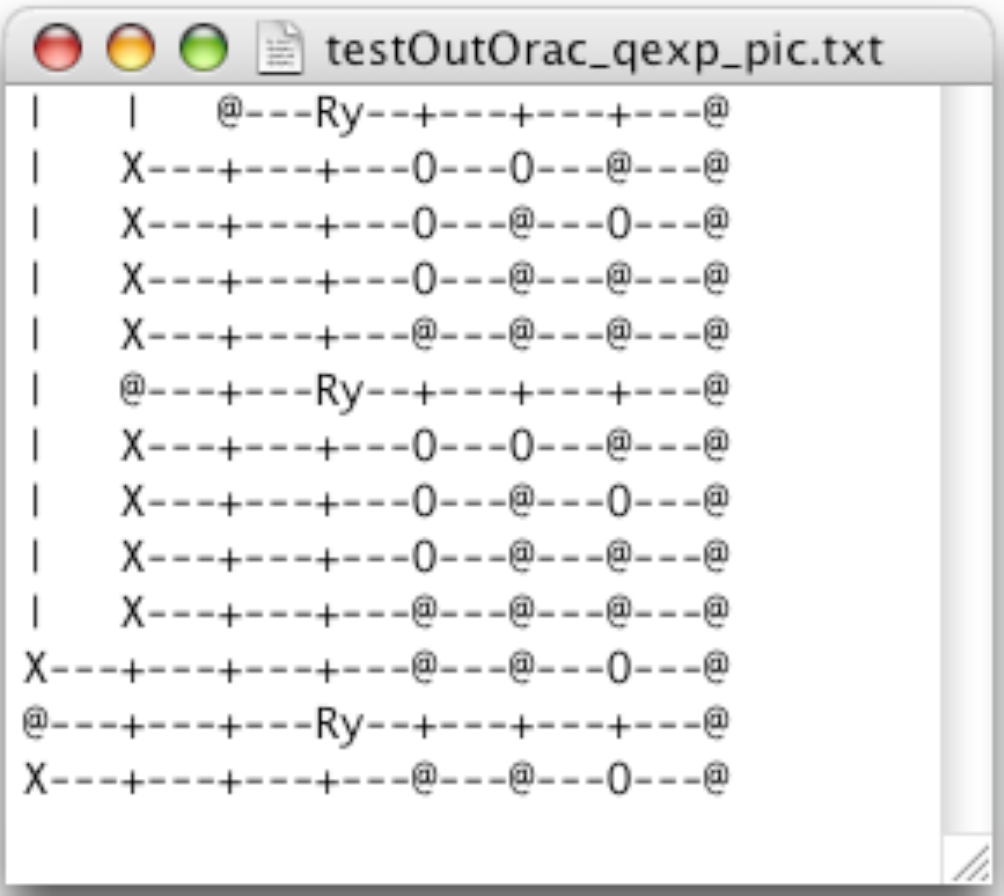}
    \caption{Output Picture File generated by
    Multiplexor Expander in the same run as the
     Log File of Fig.\ref{fig-outOrac-qexp-log}}
    \label{fig-outOrac-qexp-pic}
    \end{center}
\end{figure}

Figs. \ref{fig-outOrac-qexp-log},
\ref{fig-outOrac-qexp-eng} and
\ref{fig-outOrac-qexp-pic}
are similar to Figs.
\ref{fig-outExact-qexp-log},
\ref{fig-outExact-qexp-eng} and
\ref{fig-outExact-qexp-pic} above,
except that they
were all generated
in a single run
of Multiplexor Expander,
with
\begin{itemize}
\item
the input files of Figs. \ref{fig-qexp-eng}
and \ref{fig-qexp-pic}, and
\item
the compilation mode set to {\bf Oracular Approximation}.
\end{itemize}

\newpage

\newpage

\newpage

\appendix
\section{Appendix: Classical Simulated Annealing}
\label{app-class-sim-ann}

The goal of simulated annealing, both
classical and quantum, is to find the
minimum of a (bounded below) function (i.e.
to solve a {\bf minimization problem}). The
function $E:S_\rvx\rarrow \RR^{\geq 0}$
to be minimized will be called the {\bf energy
function} (assumed non-negative
without loss of generality). It's
domain $S_\rvx$ will be called the {\bf
state space}. We
will assume that
$S_\rvx=\{0,1,2,\ldots, \ns-1\}$,
were $\ns=2^\nb$ is the number of states and $\nb$
is the number of bits.
Besides the state space and energy
function, it is also
convenient to introduce a {\bf neighborhood
function}
$neig:S_\rvx\times S_\rvx\rarrow Bool$,
defined by
$neig(x,y)=\theta(\mbox{$x$ and $y$ are neighbors})$.

To solve this minimization problem,
classical simulated annealing uses a
Markov chain.

A {\bf Markov chain}
is a Bayesian network
$\rvx_0\rarrow \rvx_1 \rarrow
\cdots \rarrow \rvx_{t_f}$,
wherein all random variables $\rvx_j\in S_{\rvx_j}$
have the same range of values:
$S_{\rvx_j}=S_\rvx$ for all $j$,
and every
node except the first one has the same transition
probability matrix:
$P_{\rvx_{j+1}|\rvx_j}(y|x) = M(y|x)$
for $j=0,1,\ldots, t_f-1$ and $x,y\in S_\rvx$.
A {\bf stationary state} $\pi()$ of the Markov chain
with transition probability matrix $M$
is a probability distribution on $S_\rvx$ which is
also
an eigenstate of $M$ with unit eigenvalue,

\beq
\sum_{x\in S_\rvx} M(y|x)\pi(x) = \pi(y)
\;
\eeq
for all $y\in S_\rvx$.
We say that a probability distribution
$\pi(x)$ is a
{\bf detailed balance} of $M$ if
\beq
M(y|x)\pi(x) = M(x|y)\pi(y)
\;
\label{eq-detBal}
\eeq
for all $x,y\in S_\rvx$.
Clearly, if $\pi()$
is a detailed balance of $M$,
it is also a stationary state of it.

\begin{figure}[h]
    \begin{center}
    \includegraphics[height=1.5in]{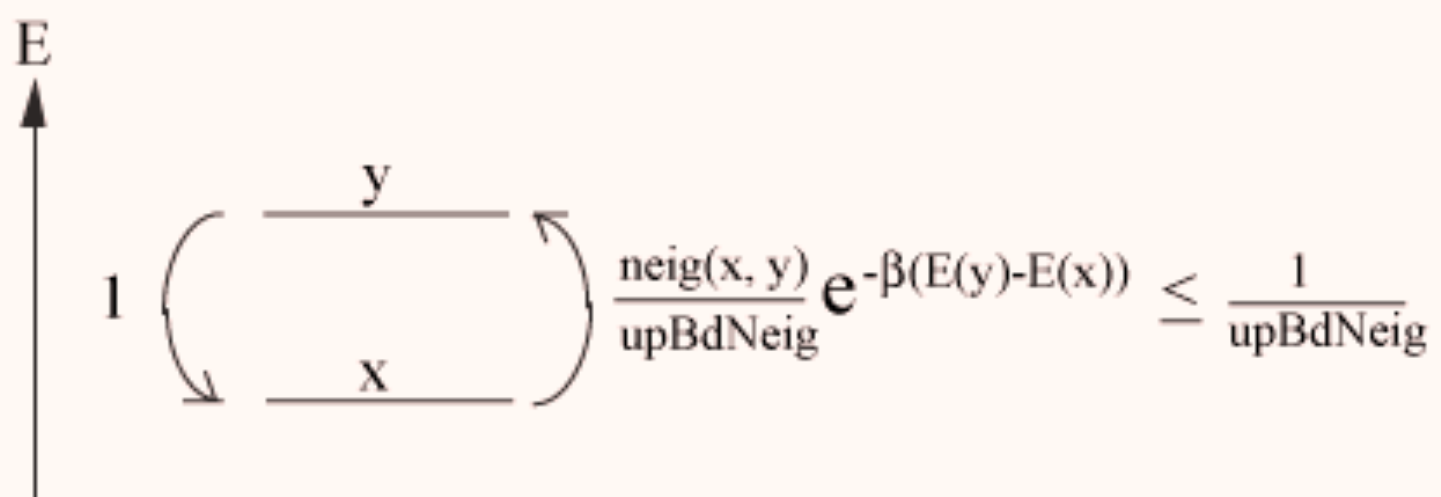}
    \caption{Transition probabilities
    for Metropolis Algorithm.}
    \label{fig-metropolis}
    \end{center}
\end{figure}

Classical
simulated annealing uses
a special Markov chain due to Metropolis.
The Metropolis transition probability matrix $M_\beta$
for a given minimization problem
and inverse temperature $\beta>0$,
is defined as follows:

\beq
M_\beta(y|x)=\left\{
\begin{array}{l}
\theta(x\neq y) \frac{neig(x,y)}{upBdNeig}
\min\{ 1 , e^{-\beta[E(y)-E(x)]}\}
\\
+ \theta(x=y)\left(1 - \sum_{z,z\neq x} M_\beta(z|x)\right)
\end{array}
\right.
\;,
\label{eq-metro}
\eeq
where
$upBdNeig$ is some real number
greater or equal to $\max_y \sum_x neig(x,y)$.
Thus, $upBdNeig$
 is an upper bound on the number of neighbors.
Fig.\ref{fig-metropolis}
tries to explain the logic behind
Eq.(\ref{eq-metro}). A ``system"
prefers going downhill to going uphill,
but is willing to visit a neighbor
living uphill occasionally.

One can show that the following
probability distribution (called a
{\bf normalized Boltzmann factor})
is a detailed balance and therefore a
stationary distribution of $M_\beta$.

\beq
\pi_\beta(x)=\frac{e^{-\beta E(x)}}{Z_\beta}
\;.
\label{eq-boltz-fac}
\eeq
$Z_\beta$ is called the
{\bf partition function}. It is defined so as
to make $\pi()$ a probability distribution:

\beq
Z_\beta = \sum_x e^{-\beta E(x)}
\;.
\eeq

In classical simulated annealing,
we consider a product
$M_{\beta_{t_f}} \ldots M_{\beta_1} M_{\beta_0}$
of
transition probability matrices.
The monotonically
increasing (or at least non-decreasing)
sequence
of non-negative real numbers $\beta_0,\beta_1 \ldots, \beta_{t_f}$
is called the {\bf annealing schedule}.

\section{Appendix: Q-Embedding of
Probability Matrix}\label{app-q-emb}

In this Appendix, we will
review the concept of
q-embedding of a probability matrix,
as used in Ref.\cite{TucInfer} and later
in Ref.\cite{TucMetHas}.

Given a conditional probability $P(y|x)$
where $x\in S_\rvx$ and $y\in S_\rvy$,
we will call a {\bf probability matrix}
the matrix $P$
with entries $P(y|x)$
with rows labeled by the $y$ and
columns by the $x$.
Any
unitary matrix $U$
with matrix elements
$\bra{y}\bra{\tilde{x}}U\ket{\tilde{y}}\ket{x}=
A(y, \tilde{x}|\tilde{y}, x)$,
where $x,\tilde{x}\in S_\rvx$ and
$y, \tilde{y}\in S_\rvy$, that
satisfies

\beq
\sum_{\tilde{x}}|A(y, \tilde{x}|\tilde{y}=0, x)|^2=
P(y|x)
\;,
\label{eq-q-emb-gen-summed}
\eeq
for all $x,y$, will be called a {\bf q-embedding
(quantum-embedding)
of the probability matrix} $P$.
$A()$ acts like a probability amplitude.
The index $\tilde{y}$ that we set to
a fixed value (call it zero)
is called a source
and the index $\tilde{x}$
that we sum over is called a sink.
This nomenclature is similar to
the one used by Fredkin and Toffoli
when they showed how any boolean
function $f:Bool^m\rarrow Bool^n$
can be embedded in a
reversible function. See Ref.\cite{TucInfer}
for more details and references.
Note that Eq.(\ref{eq-q-emb-gen-summed})
 is satisfied if we set

\beq
A(y, \tilde{x}|\tilde{y}=0, x)=
\delta_x^{\tilde{x}}
\sqrt{P(y|x)}
\;.
\eeq

A q-embedding of a probability
matrix is of course not unique. Next
we will give one possible q-embedding
for any square probability
matrix acting on $\nb$ bits.
The q-embedding
that we will give
is very convenient because it is
expressed succinctly
as a product of quantum
multiplexors. (See Ref.\cite{notation}
for a review of quantum multiplexors).

We begin by pointing out some trivial
algebraic results that will be used
below.
Note that for any $\theta\in \RR$,

\beq
e^{-i\sigy\theta}
=
\left[
\begin{array}{cc}
C_\theta&-S_\theta\\
S_\theta & C_\theta
\end{array}
\right]
\;.
\eeq
$C_\theta, S_\theta$ are
shorthand for $\cos\theta$ and $\sin\theta$,
respectively.
Thus, if $b\in Bool$,
we get from the
first column of
this matrix that:

\beq
\bra{b}e^{-i\sigy\theta}\ket{0}
=
C^{\bar{b}}_\theta
S^b_\theta
=
\left\{
\begin{array}{l}
C_\theta\;\;{\rm if}\;\;b=0\\
S_\theta\;\;{\rm if}\;\;b=1
\end{array}
\right.
\;.
\label{eq-rot-mat-elem}
\eeq
(We are using the notation of Ref.\cite{notation}
for $\overline{b}$ where $b\in Bool$, namely $\overline{0}=1$
and $\overline{1}=0$.)
A compact expression for the entries of
both columns can be obtained
as follows. For $a,b\in Bool$,

\beq
\bra{b}e^{-i\sigy \theta}\ket{a}
=
\bra{b}e^{-i\sigy \theta}\sigx^a\ket{0}
=
\bra{b\oplus a}e^{-i(-1)^a \sigy\theta}\ket{0}
\;.
\label{eq-gen-rot-mat-elem}
\eeq
The last matrix element
in Eq.(\ref{eq-gen-rot-mat-elem}) is given
by Eq.(\ref{eq-rot-mat-elem}).

Let $q\in \RR^{\ns\times\ns}$ be the
probability matrix for which we desire a
q-embedding.
First consider $\nb=1$.
We want
the following constraint to be satisfied:

\beq
A(b_1, b_0|a_1=0, a_0)=
\delta_{b_0}^{a_0}\sqrt{q(b_1|a_0)}
\;,
\eeq
where $b_1,b_0,a_1,a_0\in Bool$.
Define a unitary matrix $\check{U}$ by

\beq
\check{U}=\;\;\;
\begin{array}{c}
\Qcircuit @C=1em @R=1em @!R{
\lstick{\mbox{\tiny{0}}}
&\muxorgate
&\qw
\\
\lstick{\mbox{\tiny{1}}}
&\emptygate\qwx
&\qw
}
\end{array}
=
e^{-i\sum_{a_0}\sigy(1)\theta_{|a_0}P_{a_0}(0)}
\;,
\eeq
where the angles $\theta_{|a_0}$
are defined by

\beq
C_{\theta_{|a_0}}^{2\bar{b}_1}
S_{\theta_{|a_0}}^{2b_1}
=q_{b_1|a_0}
\;.
\eeq
Using $P_b\ket{b'}=\delta_b^{b'}\ket{b}$ for $b,b'\in Bool$,
and Eq.\ref{eq-rot-mat-elem},
it follows that

\beqa
\begin{array}{c}
\Qcircuit @C=1em @R=1em @!R{
\bra{b_0}\;\;\;\;\;
&\muxorgate
&\qw\;\;\;\;\;\;\ket{a_0}
\\
\bra{b_1}\;\;\;\;\;
&\emptygate\qwx
&\qw\;\;\;\;\;\;\ket{0}
}
\end{array}
\;\;\;\;&=&
\delta_{b_0}^{a_0}
\bra{b_1}
e^{-i\sigy(1)\theta_{|a_0}}
\ket{0}
\\
&=&
\delta_{b_0}^{a_0}
C_{\theta_{|a_0}}^{\bar{b}_1}
S_{\theta_{|a_0}}^{b_1}
\\
&=&
\delta_{b_0}^{a_0}\sqrt{q_{b_1|a_0}}
\;.
\eeqa
Hence $\check{U}$ is a q-embedding of
$q$.

Now consider $\nb=2$. We want
\beq
A(b_3, b_2, b_1, b_0|a_3=0, a_2=0, a_1, a_0)=
\delta_{b_1}^{a_1}
\delta_{b_0}^{a_0}\sqrt{q(b_3, b_2|a_1, a_0)}
\;,
\eeq
where all $a_j$ and $b_j$ are in $Bool$.
Define a unitary matrix $\check{U}$ by

\beq
\check{U}=\;\;\;
\begin{array}{c}
\Qcircuit @C=1em @R=1em @!R{
\lstick{\mbox{\tiny{0}}}
&\muxorgate
&\muxorgate
&\qw
\\
\lstick{\mbox{\tiny{1}}}
&\muxorgate
&\muxorgate
&\qw
\\
\lstick{\mbox{\tiny{2}}}
&\muxorgate
&\emptygate\qwx[-2]
&\qw
\\
\lstick{\mbox{\tiny{3}}}
&\emptygate\qwx[-3]
&\qw
&\qw
}
\end{array}
=
e^{-i\sum_{b_2,a_1,a_0}\sigy(3)\theta_{b_2|a_1a_0}P_{b_2 a_1 a_0}(2,1,0)}
e^{-i\sum_{a_1,a_0}\sigy(2)\theta_{|a_1a_0}P_{a_1 a_0}(1,0)}
\;,
\eeq
where the angles $\theta_{b_2|a_1a_0}$
and $\theta_{|a_1a_0}$
are defined by

\beq
C_{\theta_{b_2|a_1a_0}}^{2\bar{b}_3}
S_{\theta_{|a_1a_0}}^{2b_3}
=\frac{q_{b_3b_2|a_1a_0}}
{q_{.\; b_2|a_1a_0}}
\;,
\eeq
and

\beq
C_{\theta_{|a_1a_0}}^{2\bar{b}_2}
S_{\theta_{|a_1a_0}}^{2b_2}
=q_{.\; b_2|a_1a_0}
\;.
\eeq
(We are using the notation of Ref.\cite{notation}
where a dot at the position of
an index means that the index has
been summed over; e.g., $q_{.\;b}=\sum_a q_{ab}$).
It follows that

\beqa
\begin{array}{c}
\Qcircuit @C=1em @R=1em @!R{
\bra{b_0}\;\;\;\;\;
&\muxorgate
&\muxorgate
&\qw\;\;\;\;\;\;\ket{a_0}
\\
\bra{b_1}\;\;\;\;\;
&\muxorgate
&\muxorgate
&\qw\;\;\;\;\;\;\ket{a_1}
\\
\bra{b_2}\;\;\;\;\;
&\muxorgate
&\emptygate\qwx[-2]
&\qw\;\;\;\;\;\;\ket{0}
\\
\bra{b_3}\;\;\;\;\;
&\emptygate\qwx[-3]
&\qw
&\qw\;\;\;\;\;\;\ket{0}
}
\end{array}
\;\;\;\;&=&
\delta_{b_1}^{a_1}
\delta_{b_0}^{a_0}
\bra{b_3}
e^{-i\sigy(3)\theta_{b_2|a_1a_0}}
\ket{0}
\bra{b_2}
e^{-i\sigy(2)\theta_{|a_1a_0}}
\ket{0}
\\
&=&
\delta_{b_1}^{a_1}
\delta_{b_0}^{a_0}
\left[C_{\theta_{b_2|a_1a_0}}^{\bar{b}_3}
S_{\theta_{b_2|a_1a_0}}^{b_3}\right]
\left[C_{\theta_{|a_1a_0}}^{\bar{b}_2}
S_{\theta_{|a_1a_0}}^{b_2}\right]
\\
&=&
\delta_{b_1}^{a_1}
\delta_{b_0}^{a_0}
\sqrt{q_{b_3b_2|a_1a_0}}
\;.
\eeqa
Thus, as in the $\nb=1$ case,
$\check{U}$ is a q-embedding of
$q$.

It's clear how to generalize this
construction so
as to get a q-embedding
of a probability matrix $q\in \RR^{\ns\times\ns}$
for any positive integer $\nb$.

\section{Appendix: Szegedy Quantum \\Walk Operator $W$}
In this appendix,
we will review the
definition
and some useful
properties
of the Szegedy quantum
walk operator $W$ (first defined
by Szegedy in Ref.\cite{szegedy},
first used for simulated
annealing by Somma et al. in Ref.\cite{somma},
first implemented in terms of multiplexors here).

\subsection{Symmetric Matrix $\msym$}
For any Markov chain
with transition probability $M$,
define matrix $\Lam$ (the entry-wise
square root of $M$) by
\beq
\Lam(y|x)= \sqrt{M(y|x)}
\;,
\eeq
and the matrix $\msym$
(a symmetric version of $M$) by

\beq
\msym(y|x)= \Lam(x|y)\Lam(y|x)
\;,
\eeq
for all $x,y\in S_\rvx$.(Note that unlike $M(y|x)$,
$\msym(y|x)$ is not a probability
function in $y$, its first
argument.)

Define the quantum states

\beq
\ket{(\pi)^\eta}=
\sum_x [\pi(x)]^\eta\ket{x}
\;
\eeq
for $\eta=\frac{1}{2}, 1$. (Note that only the
$\eta=\frac{1}{2}$
state is normalized in the sense
of quantum mechanics.)

\begin{claim}

\beq
M\ket{\pi}=\ket{\pi}
\;,
\eeq
and

\beq
\msym\ket{\sqrt{\pi}}=\ket{\sqrt{\pi}}
\;.
\eeq
Also, $M$ and $\msym$ have the same
eigenvalues.
\end{claim}
\proof

Taking the square root
of both sides of the detailed balance
statement Eq.(\ref{eq-detBal}), we get

\beq
\Lam(y|x)\sqrt{\pi(x)}=
\Lam(x|y)\sqrt{\pi(y)}
\;.
\eeq
Therefore,

\beqa
\msym(y|x)=\Lam(x|y)
\frac{1}{\sqrt{\pi(x)}}
\Lam(x|y)\sqrt{\pi(y)}
&=&
\frac{1}{\sqrt{\pi(x)}}
M(x|y)
\sqrt{\pi(y)}
\;.
\eeqa
Hence,

\beq
\sum_x M(y|x)\pi(x)=
\sum_x M(x|y)\pi(y)=\pi(y)
\;,
\eeq
and

\beq
\sum_x \msym(y|x)\sqrt{\pi(x)}
=
\sum_x
\frac{1}{\sqrt{\pi(x)}}M(x|y)\sqrt{\pi(y)}
\sqrt{\pi(x)}
=\sqrt{\pi(y)}
\;.
\eeq

Order the elements of
the finite set $S_\rvx$
in some preferred way.
Use this preferred order
to
represent $M$ and $\msym$
as matrices.
Define a diagonal matrix $D$
whose diagonal entries are the
numbers $\pi(x)$ for each $x\in S_\rvx$,
with the $x$ ordered in the
preferred order:

\beq
D= diag[\left(\pi(x)\right)_{\forall x}]
\;.
\eeq
Since

\beq
\msym = D^{-\frac{1}{2}}M D^{\frac{1}{2}}
\;,
\eeq
it follows that

\beq
\det(M - \lambda)=
\det(\msym - \lambda)
\;
\eeq
for any $\lambda\in \RR$.
\qed

Let the eigenvalues\footnote{There must
be a single eigenvalue 1 and all others
must have a magnitude strictly smaller
than one because of the Frobenius-Perron Theorem.
The eigenvalues must be real (but they can be negative)
because $\msym$ is a Hermitian matrix.}
 of $\msym$ (and also of $M$)
be $m_0, m_1, \ldots m_{\ns-1}\in \RR$ with
$m_0=1\gneq |m_1|\geq
|m_2|\ldots\geq|m_{\ns-1}|$.
Define $\ket{m_j}$ to be the corresponding
eigenvectors of $\msym$ (but not necessarily
of $M$). Thus

\beq
\msym\ket{m_j} = m_j\ket{m_j}
\;,
\eeq
for $j=0,1,\ldots,\ns-1$.
In particular, $\ket{m_0}=\ket{\sqrt{\pi}}$.

For each $j$,
define $\varphi_j\in[0,\frac{\pi}{2}]$
and $\eta_j\in\{0,\pi\}$ so that
$m_j = e^{i\eta_j}\cos\varphi_j $.
(Thus, $\cos\varphi_j\geq 0$ and $e^{i\eta_j}=\pm 1$).
Note that $m_0=1$ so $\varphi_0=0$.
The $M$ eigenvalue gap $\delta$ is defined as
$\delta=1-|m_1|$. $\delta\approx\frac{\varphi_1^2}{2}$
when $\varphi_1$ is small.

\subsection{Q-Embeddings $\check{U}$ and $\hat{U}$}
Next we will consider
two ``dual" disjoint sets of distinct
qubits with $\nb$ qubits in each set.
Let these two
set be labeled
$\vec{\alpha}=(\alpha_1,\alpha_2,\ldots,\alpha_\nb)$
and
$\vec{\beta}=(\beta_1,\beta_2,\ldots,\beta_\nb)$.
Any operator $\Omega$
acting on the tensor
product of a state $\ket{x}_{\vec{\alpha}}$
and a state $\ket{y}_{\vec{\beta}}$
where $x,y\in Bool^{N_B}$,
can be represented,
depending on taste,
either in quantum
circuit notation or Dirac notation, by

\beq
\begin{array}{c}
\Qcircuit @C=1em @R=2em @!R{
&\multigate{1}{\mbox{$\Omega$}}
&\qw
&\ket{x}_{\vec{\alpha}}
\\
&\ghost{1}{\mbox{$\Omega$}}
&\qw
&\ket{y}_{\vec{\beta}}
}
\end{array}
\;\;\;\;
=
\Omega\ket{y}_{\vec{\beta}}\ket{x}_{\vec{\alpha}}
\;.
\eeq

Let $\swap$
denote the operator
that swaps all bits $\alpha_j$ and
$\beta_j$ for $j=1,2, \ldots,\nb$.

Let $\check{U}$
be any unitary matrix satisfying
for any $x\in Bool^\nb$,

\beq
\begin{array}{c}
\Qcircuit @C=1em @R=2em @!R{
&\multigate{1}{\mbox{$\check{U}$}}
&\qw
&\ket{x}
\\
&\ghost{1}{\mbox{$\check{U}$}}
&\qw
&\ket{0}
}
\end{array}
\;\;
=
\;\;
\begin{array}{c}
\Qcircuit @C=1em @R=1em @!R{
&\qw
&\qw
&\ket{x}
\\
&\gate{\Lam}
&\qw
&\ket{x}
}
\end{array}
\mbox{     or    }
\check{U}\ket{0}\ket{x}=
(\Lam\ket{x})\ket{x}
\;.
\label{eq-u0x}
\eeq
Let

\beq
\check{A}(y',y|0,x)=\;\;\;\;
\begin{array}{c}
\Qcircuit @C=1em @R=2em @!R{
\bra{y}\;\;\;\;
&\multigate{1}{\mbox{$\check{U}$}}
&\qw
&\ket{x}
\\
\bra{y'}\;\;\;\;
&\ghost{1}{\mbox{$\check{U}$}}
&\qw
&\ket{0}
}
\end{array}
=
\bra{y'}\bra{y}\check{U}\ket{0}\ket{x}
\;.
\eeq
for $x,0,y,y'\in Bool^\nb$.
Then,
by virtue of Eq.(\ref{eq-u0x}),

\beq
\check{A}(y',y|0,x)=
\delta(y,x)\Lam(y'|x)=
\delta(y,x)\sqrt{M(y'|x)}
\;.
\eeq
Thus, $\check{U}$
is a
q-embedding of
the probability matrix $M$.

If we define
$\hat{U}$  by

\beq
\hat{U} = \swap \check{U}\swap
\;,
\eeq
then
we can
immediately
write the following
equations,
which are dual
to equations
we wrote previously
for $\check{U}$:

\beq
\begin{array}{c}
\Qcircuit @C=1em @R=2em @!R{
&\multigate{1}{\mbox{$\hat{U}$}}
&\qw
&\ket{0}
\\
&\ghost{1}{\mbox{$\hat{U}$}}
&\qw
&\ket{x}
}
\end{array}
\;\;
=
\;\;
\begin{array}{c}
\Qcircuit @C=1em @R=1em @!R{
&\gate{\Lam}
&\qw
&\ket{x}
\\
&\qw
&\qw
&\ket{x}
}
\end{array}
\mbox{     or    }
\hat{U}\ket{x}\ket{0}=
\ket{x}\Lam\ket{x}
\;,
\eeq

\beq
\hat{A}(y',y|x,0)=\;\;\;\;
\begin{array}{c}
\Qcircuit @C=1em @R=2em @!R{
\bra{y}\;\;\;\;
&\multigate{1}{\mbox{$\hat{U}$}}
&\qw
&\ket{0}
\\
\bra{y'}\;\;\;\;
&\ghost{1}{\mbox{$\hat{U}$}}
&\qw
&\ket{x}
}
\end{array}
=
\bra{y'}\bra{y}\hat{U}\ket{x}\ket{0}
\;,
\eeq

\beq
\hat{A}(y',y|x,0)=
\delta(y',x)\Lam(y|x)=
\delta(y',x)\sqrt{M(y|x)}
\;.
\eeq

Next define the unitary
operator $U$ by

\beq
U = \hat{U}^\dagger \check{U}
\;.
\label{eq-u-uhat-ucheck}
\eeq
Clearly,

\beq
\swap U \swap = U^\dagger
\;.
\eeq
Note that $U\swap$ is
Hermitian and its square equals one:

\beq
(U\swap)^\dagger
=
\swap U^\dagger
=
U\swap
\;\;,
\;\;
(U\swap)^2 = 1
\;.
\eeq
Matrix $U$ has the
following highly desirable property:

\begin{claim}
For any $j,k\in\{0,1,\dots,N_S-1\}$,

\beq
\begin{array}{r}
\bra{0}
\\
\bra{m_j}
\end{array}
U
\begin{array}{l}
\ket{m_k}
\\
\ket{0}
\end{array}
=
m_j \delta_j^k
\;.
\eeq
\end{claim}
\proof
\beqa
\begin{array}{r}
\bra{0}
\\
\bra{m_j}
\end{array}
\hat{U}^\dagger \check{U}
\begin{array}{l}
\ket{m_k}
\\
\ket{0}
\end{array}
&=&
\sum_{y,x}
\begin{array}{r}
\;
\\
\av{m_j|y}
\end{array}
\left[
\begin{array}{l}
\bra{y}\Lam^T
\\
\bra{y}
\end{array}
\right]\left[
\begin{array}{r}
\ket{x}
\\
\Lam\ket{x}
\end{array}
\right]
\begin{array}{l}
\av{x|m_k}
\\
\;
\end{array}
\\
&=&
\sum_{y,x}
\av{m_j|y}\Lam^T(y|x)\Lam(y|x)\av{x|m_k}
\\
&=&
\av{m_j|\msym|m_k}=
m_j \delta_j^k
\;.
\eeqa
\qed

\subsection{Definition of $W$}

Now define the projection operator
$\hat{\pi}$ by
(expressed below in 3 alternative but equivalent notations)

\beq
\hat{\pi}=
\begin{array}{c}
\Qcircuit @C=1em @R=.25em @!R{
&\gate{\ket{0}\bra{0}}
&\qw
\\
&\qw
&\qw
}
\end{array}
=
I_2^{\otimes\nb}\otimes P_0^{\otimes\nb}
=
P_0(\vec{\alpha})
\;.
\eeq
Define the projection operator $\check{\pi}$
dual to $\hat{\pi}$, by

\beq
\check{\pi} =
\swap\hat{\pi}\swap
=
\begin{array}{c}
\Qcircuit @C=1em @R=.25em @!R{
&\qw
&\qw
\\
&\gate{\ket{0}\bra{0}}
&\qw
}
\end{array}
\;.
\eeq
Define a reflection
operator $(-1)^{\hat{\pi}}$
(expressed below in several equivalent notations)
and its dual reflection
operator $(-1)^{\check{\pi}}$
as follows:

\beq
(-1)^{\hat{\pi}}
=
1-2\hat{\pi}
=
 I_2^{\otimes\nb}\otimes (-1)^{P_0^{\otimes\nb}}
=
(-1)^{\prod_{j=1}^{\nb}\nbar(\alpha_j)}
\;,
\eeq

\beq
(-1)^{\check{\pi}}
=
\swap
(-1)^{\hat{\pi}}
\swap
\;.
\eeq

Finally (gasp!), we are ready
to define the {\bf Szegedy quantum walk
operator} $W$ corresponding to the
transition probability matrix $M$, by

\beq
W(M)=U (-1)^{\check{\pi}} U^\dagger
(-1)^{\hat{\pi}}
\;.
\eeq

\subsection{Eigenvalues of $W$}

To find the eigenvalues of $W$,
we will use the following identities.

\begin{claim}\label{cl-proj-ids}
\begin{subequations}
\beq
\hat{\pi}\ket{m_j 0} = \ket{m_j 0}
\;,
\eeq

\beq
\hat{\pi}(U\swap)\ket{m_j 0} = m_j\ket{m_j 0}
\;,\label{eq-pi-u-swap}
\eeq
\end{subequations}
for all $j\in \{0,1,\ldots,N_S-1\}$.
\end{claim}
\proof

From the definition of $\hat{\pi}$,
we see that
\beq
\hat{\pi}
\begin{array}{l}
\ket{0}
\\
\ket{m_j}
\end{array}
=
\begin{array}{l}
\ket{0}
\\
\ket{m_j}
\end{array}
\;.
\eeq
Also,

\beq
\hat{\pi}
(U\swap)
\begin{array}{l}
\ket{0}
\\
\ket{m_j}
\end{array}
=
\sum_k
\begin{array}{r}
\ket{0}\bra{0}
\\
\ket{m_k}\bra{m_k}
\end{array}
U
\begin{array}{l}
\ket{m_j}
\\
\ket{0}
\end{array}
=
m_j
\begin{array}{l}
\ket{0}
\\
\ket{m_j}
\end{array}
\;.
\eeq
\qed

An immediate
consequence of Claim \ref{cl-proj-ids} is that

\beq
\bra{m_{j} 0}
U\swap \ket{m_k 0}
=
\bra{m_{j} 0}\hat{\pi}
U\swap \ket{m_k 0}
=
m_j\delta_j^{k}
\;,
\label{eq-u-swap-mat-elem}
\eeq
for $j,k\in\{0,1,\ldots,\ns-1\}$.

Note that since $m_0=1$,
Eq.(\ref{eq-u-swap-mat-elem}) implies that

\beq
\ket{m_00} = U\swap\ket{m_00}
\;.
\label{eq-u-swap-equals-1}
\eeq

Another consequence
of Claim \ref{cl-proj-ids}
is that $\ket{m_0 0}$
is a stationary state of $W$. Indeed,
one has

\beqa
W\ket{m_0 0}&=&
U
(-1)^{\check{\pi}}
U^\dagger
(-1)^{\hat{\pi}}
\ket{m_0 0}
\\
&=&
U
\swap
(1-2\hat{\pi})
\swap
U^\dagger
(-1)
\ket{m_0 0}
\\
&=&
(1-2m_0U\swap)(-1)
\ket{m_0 0}
\\
&=&
(1-2)(-1)
\ket{m_0 0}
\\
&=&\ket{m_0 0}
\;.
\eeqa

Let
\beq
\calv_{busy}^j=span\{\ket{m_j0}, U\swap\ket{m_j0}\}
\;
\eeq
for $j\in\{0,1,\ldots,N_S-1\}$.
(By ``span" we mean all
linear combinations of
these vectors with {\it complex}
coefficients.)

\begin{claim}\label{cl-v-busy-j}
$W\calv_{busy}^j=\calv_{busy}^j$
for all $j\in\{0,1,\ldots,N_S-1\}$.
For $j=0$, let

\beq
\ket{\psi_0} = \ket{m_0 0}
\;.
\eeq
$\{\ket{\psi_0}\}$ is an
orthonormal basis for $\calv^0_{busy}$
and $W\ket{\psi_0} = \ket{\psi_0}$.
For $j\neq 0$, let

\beq
\ket{\psi_{\pm j}}=
\frac{\pm i}{\sqrt{2} \sin\varphi_j}
(e^{-i\eta_j}U\swap\ket{m_j0}- e^{\pm i2\varphi_j}\ket{m_j0})
\;.
\label{eq-psi-j-original-basis}
\eeq
$\{\ket{\psi_j}, \ket{\psi_{-j}}\}$ is an
orthonormal basis for $\calv^j_{busy}$
and $W\ket{\psi_{\pm j}} =
e^{\pm i2\varphi_j}\ket{\psi_{\pm j}}$.
\end{claim}
\proof

Using the identities of
Claim \ref{cl-proj-ids},
one finds after some algebra that

\begin{subequations}
\label{eq-w-invariant-plane}
\beq
W\ket{m_j 0}=
(-1)\ket{m_j 0}
+
(2m_j)U\swap \ket{m_j 0}
\;,
\eeq
and

\beq
W(U\swap)\ket{m_j 0}=
(-2m_j)\ket{m_j 0}
+
(-1 + 4m_j^2)U\swap \ket{m_j 0}
\;
\eeq
\end{subequations}
for all $j$.

According
to Eqs.(\ref{eq-w-invariant-plane}),
$\calv_{busy}^j$ is invariant under
the action of $W$ for each $j$.
By virtue of Eq.(\ref{eq-u-swap-mat-elem}),
$\calv_{busy}^j$
is 1-dim for $j=0$ and 2-dim if $j\neq 0$.
We've already proven that
$\ket{m_0 0}$
is a stationary state of $W$.

\begin{figure}[h]
    \begin{center}
    \includegraphics[scale=.50]{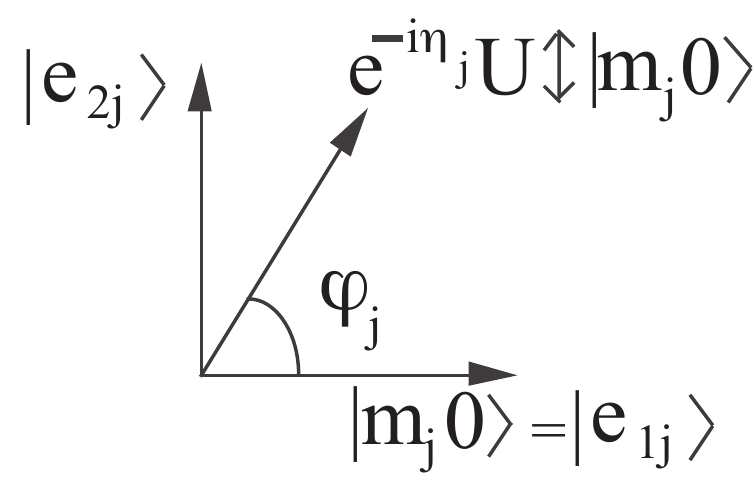}
    \caption{Definition of $\ket{e_{1j}}$
    and $\ket{e_{2j}}$.}
    \label{fig-e1-e2-def}
    \end{center}
\end{figure}

Now consider the case $j\neq0$.
Both $U(-1)^{\check{\pi}}U^\dagger$ and
$(-1)^{\hat{\pi}}$ are
reflections in the planar subspace $\calv_{busy}^j$,
and reflections are a special type
of rotation about the axis normal
to this plane, so their product is
also a rotation about this axis.
The vectors $\ket{m_j0}$,
and $U\swap\ket{m_j0}\}$ are
independent but not orthogonal.
However, we can express them in terms of orthogonal vectors
(see Fig.\ref{fig-e1-e2-def}) as follows:

\begin{subequations}
\beq
\ket{m_j0} = \ket{e_{1j}}
\;,
\eeq
and

\beq
e^{-i\eta_j}U\swap\ket{m_j0} =
\cos(\varphi_j)\ket{e_{1j}}+
\sin(\varphi_j)\ket{e_{2j}}
\;.
\eeq
\end{subequations}
In the $\ket{e_{1j}}$, $\ket{e_{2j}}$ basis,
we find after substituting
$m_j = e^{i\eta_j}\cos(\varphi_j)$ into
Eqs.(\ref{eq-w-invariant-plane}) that

\beq
W=
\left[
\begin{array}{cc}
\cos(2\varphi_j)
&\sin(2\varphi_j)\\
-\sin(2\varphi_j)
&\cos(2\varphi_j)
\end{array}
\right]
\;.
\eeq
The eigenvalues of this matrix
are $e^{\pm i2\varphi_j}$, with
corresponding eigenvectors:

\beq
\ket{\psi_{\pm j}} =
\frac{1}{\sqrt{2}}(\ket{e_{1j}}\pm \ket{e_{2j}})
\;.
\label{eq-psi-j-e-basis}
\eeq
These eigenvectors satisfy

\beq
\av{\psi_{-j}|\psi_{j}}=0\;\;,\;\;
\av{\psi_{j}|\psi_{j}}=1
\;.
\eeq
By expressing $\ket{e_{1j}}$
and $\ket{e_{2j}}$ in
Eq.(\ref{eq-psi-j-e-basis})
in the original basis,
we get Eq.(\ref{eq-psi-j-original-basis}).
\qed

Define the following vector spaces:

\beq
\calv= span\{ \ket{x}\otimes\ket{y}: x,y\in S_\rvx\}
\;,
\eeq

\beq
\calv_A= span\{ \ket{x}\otimes\ket{0}: x\in S_\rvx\}
\;,
\eeq

\beq
\calv_B = U\swap \calv_A
\;,
\eeq
and

\beq
\calv_{busy} = \calv_A + \calv_B
\;.
\eeq
$\calv$ can be expressed as
a direct sum of
$\calv_{busy}$ and its orthogonal
complement $\calv_{busy}^\perp$:

\beq
\calv=\calv_{busy}\oplus \calv_{busy}^\perp
\;.
\eeq
From Claim \ref{cl-v-busy-j},
it follows that
$\calv_{busy}$
is a direct sum of the subspaces $\calv_{busy}^j$:

\beq
\calv_{busy} =\bigoplus_{j=0}^{N_S-1} \calv_{busy}^j
\;.
\label{eq-v-busy-direct-sum}
\eeq
Recall that matrices
$M$ and $\msym$ are $N_S$ dimensional
whereas $W$ is $N_S^2$ dimensional.
Since
the size of $S_\rvx$
is $N_S$,
$dim(\calv) = N_S^2$.
From Eq.(\ref{eq-v-busy-direct-sum}) and
Claim \ref{cl-v-busy-j},
$dim(\calv_{busy})=2N_S-1$.
Furthermore,
$\{\ket{\psi_j}: j=0, \pm 1, \pm 2,
\ldots, \pm(\ns-1)\}$
is an orthonormal basis for $\calv_{busy}$.

At this point we've
explained the action
of $W$ on $\calv_{busy}$,
but we haven't said anything
about the action of $W$ on
$\calv^\perp_{busy}$. Next we
show that $W$
acts simply as the identity
on $\calv^\perp_{busy}$.
(This is what one would expect since
the vectors in
$\calv^\perp_{busy}$ are parallel
to the axis of the $W$ rotation.)
Recall that if $S$ and $T$
are subspaces of a vector space $\calv$,
then
$(S+T)^\perp = S^\perp\cap T^\perp$.
Therefore,

\beq
\calv^\perp_{busy} = \calv_A^\perp\cap \calv_B^\perp
\;.
\eeq
From the definitions
of $\calv_A$ and $\calv_B$,
it's easy to see that

\beq
\calv_A^\perp=span\{\ket{x}\otimes\ket{y}:
x\in S_\rvx,
\mbox{ and } y\in S_\rvx-\{0\}\}
\;,
\eeq
and

\beq
\calv_B^\perp=U\swap(\calv_A^\perp)
\;.
\eeq

\begin{claim}
\beq
W\ket{\phi} = \ket{\phi}
\;
\eeq
for all $\ket{\phi}\in \calv_{busy}^\perp$.
\end{claim}
\proof
Let $\ket{\phi}\in \calv_{busy}^\perp=
\calv_A^\perp \cap \calv_B^\perp$. Hence
$\ket{\phi}\in \calv_A^\perp$
and $\ket{\phi}=U\swap \ket{\theta}$
for some $\ket{\theta}\in \calv_A^\perp$.

\beqa
U
(-1)^{\check{\pi}}
U^\dagger
(-1)^{\hat{\pi}}
\ket{\phi}
&=&
U
\swap
(-1)^{\hat{\pi}}
\swap
U^\dagger
(-1)^0
\ket{\phi}
\\
&=&
U
\swap
(1-2\hat{\pi})
\swap
U^\dagger
U\swap\ket{\theta}
\\
&=&
U
\swap
(1-2\hat{\pi})
\ket{\theta}
\\
&=&
\ket{\phi}
\;.
\eeqa
\qed

\subsection{Multiplexor Implementation of $W$}

Consider the case $\nb=2$.
Using $e^{i\pi}=-1$ and $\ket{0}\bra{0}=\nbar$,
one gets

\beq
W = U(-1)^{\check{\pi}}U^\dagger (-1)^{\hat{\pi}}=
\begin{array}{c}
\Qcircuit @C=1em @R=1em @!R{
&\multigate{3}{U}
&\qw
&\multigate{3}{\mbox{$U^\dagger$}}
&\ogate
&\qw
\\
&\ghost{3}{U}
&\gate{e^{i\pi}}
&\ghost{3}{\mbox{$U^\dagger$}}
&\ogate\qwx[-1]
&\qw
\\
&\ghost{3}{U}
&\ogate\qwx
&\ghost{3}{\mbox{$U^\dagger$}}
&\gate{e^{i\pi}}\qwx[-1]
&\qw
\\
&\ghost{3}{U}
&\ogate\qwx
&\ghost{3}{\mbox{$U^\dagger$}}
&\qw
&\qw
}
\end{array}
\;.
\label{eq-w-nb2}
\eeq
Now we need to find a SEO for the $U$
in
Eq.(\ref{eq-w-nb2}).
Using Eq.(\ref{eq-u-uhat-ucheck})
to express $U$
 in terms
of $\check{U}$,
and using
the method given in
Appendix \ref{app-q-emb}
for implementing
$\check{U}$
in terms of multiplexors,
we get

\beq
U = \swap\check{U}^\dagger\swap\check{U}=
\begin{array}{c}
\Qcircuit @C=1em @R=1em @!R{
&\uarrowgate
&\qw
&\muxorgate
&\muxorgate
&\uarrowgate
&\qw
&\muxorgate
&\muxorgate
&\qw
\\
&\qw
&\uarrowgate
&\muxorgate
&\muxorgate
&\qw
&\uarrowgate
&\muxorgate
&\muxorgate
&\qw
\\
&\darrowgate\qwx[-2]
&\qw
&\gate{\;^\dagger}\qwx[-2]
&\muxorgate
&\darrowgate\qwx[-2]
&\qw
&\muxorgate
&\emptygate\qwx[-2]
&\qw
\\
&\qw
&\darrowgate\qwx[-2]
&\qw
&\gate{\;^\dagger}\qwx[-3]
&\qw
&\darrowgate\qwx[-2]
&\emptygate\qwx[-3]
&\qw
&\qw
}
\end{array}
\;.
\label{eq-u-mplexors}
\eeq
In Eq.(\ref{eq-u-mplexors}),
a box with a dagger in it
represents the Hermitian conjugate
of the box without a dagger and
acting earlier on the same qubit.
It's clear how to
generalize this construction
of $W$ to any number $\nb$
of bits.

\begin{figure}[h]
    \begin{center}
    \includegraphics[scale=.70]{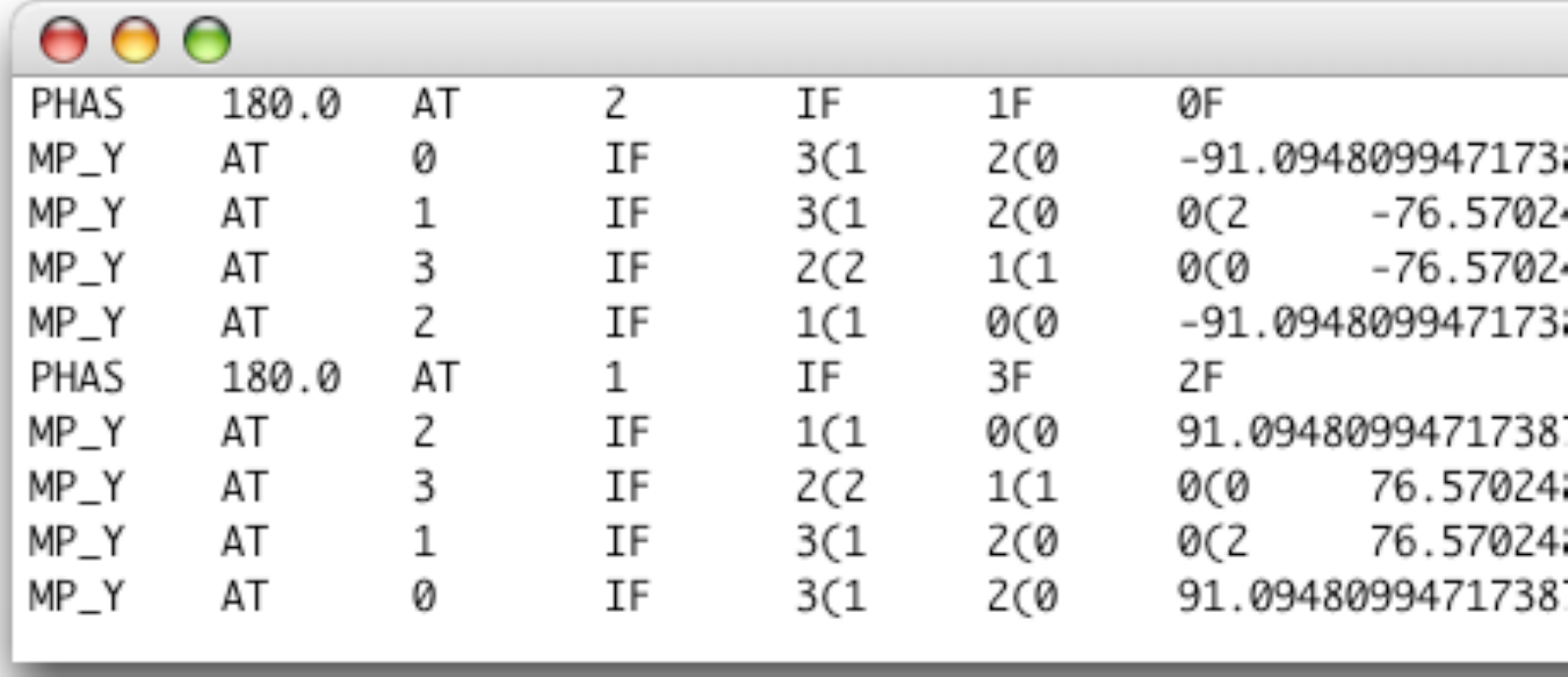}
    \caption{
    English File specifying a Szegedy quantum
    walk operator $W(M)$ for an
    $M$ with $\nb=2$.
     Right hand side of file is not visible.}
    \label{fig-sze-eng}
    \end{center}
\end{figure}

\begin{figure}[h]
    \begin{center}
    \includegraphics[scale=.70]{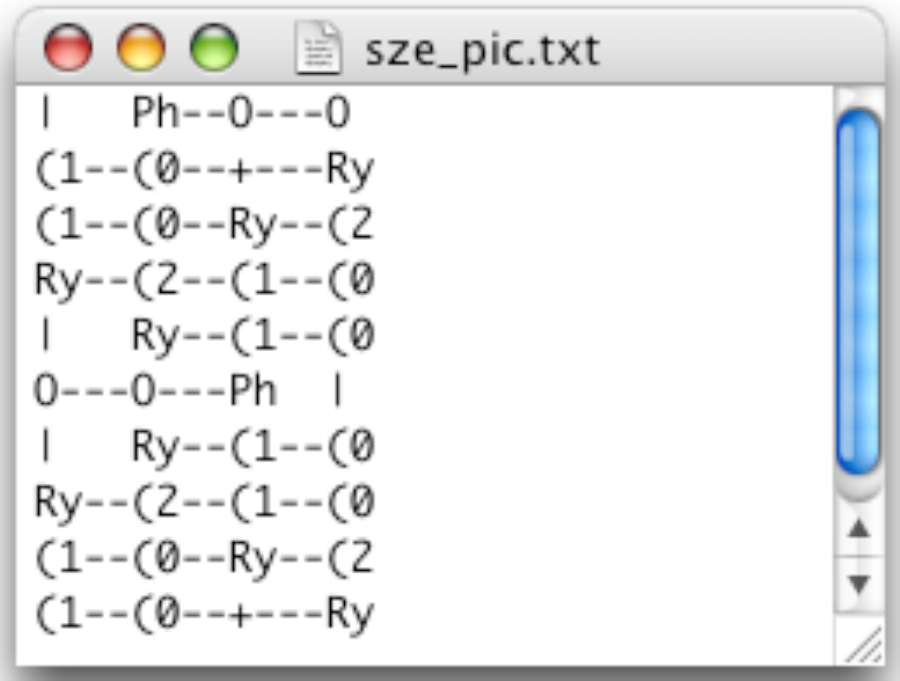}
    \caption{Picture File corresponding
    to the English File of Fig.\ref{fig-sze-eng}}
    \label{fig-sze-pic}
    \end{center}
\end{figure}

Figs.\ref{fig-sze-eng}
and \ref{fig-sze-pic}
show English and Picture
files, written using
the format of QuSAnn and
Multiplexor Expander, for a
Szegedy quantum walk operator
$W$, for a case with $\nb=2$.


\section{Appendix: Wocjan-Abeyesinghe Algorithm}
\label{app-wa-algo}
In this appendix,
we will review the
Wocjan-Abeyesinghe Algorithm
for quantum simulated annealing,
for which QuSAnn generates a
quantum circuit. This appendix follows closely
Ref.\cite{wocjan1}.

\begin{figure}[h]
    \begin{center}
    \includegraphics[scale=.60]{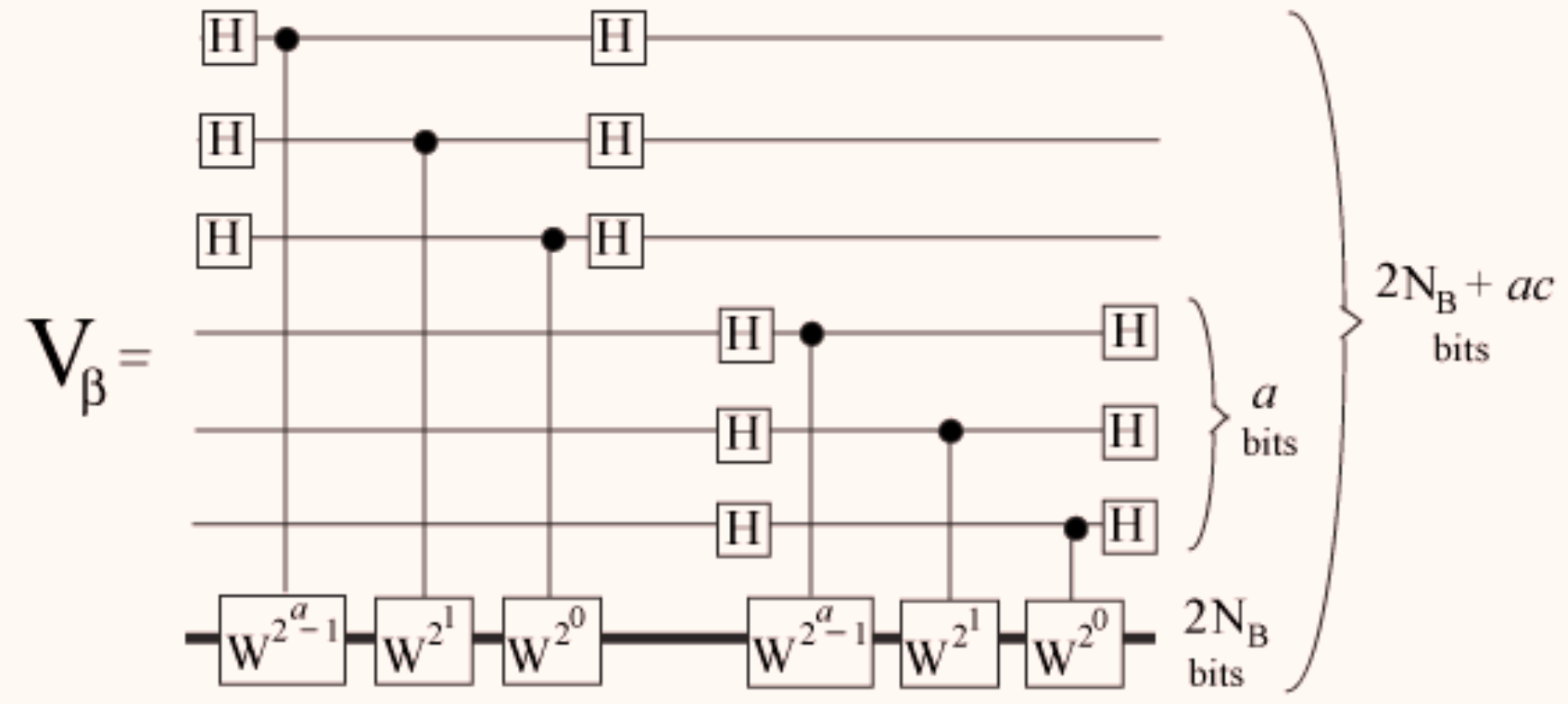}
    \caption{Definition of operator $V_\beta$
    used in Ref.\cite{wocjan1}.}
    \label{fig-v-beta}
    \end{center}
\end{figure}

For any inverse temperature $\beta>0$,
define $V_\beta$ by the
quantum circuit of Fig.\ref{fig-v-beta}.
Fig.\ref{fig-v-beta}
also defines the parameters
$a=1,2,\ldots$ which we refer
to as the {\bf number of probe bits},
and $c=1,2,\ldots$ which
we refer to as the {\bf number of
phase estimation (PE) steps}.

Let
\beq
Q = (e^{i\frac{\pi}{3}})^{P_0^{\otimes a c}}
\;
\eeq
and

\beq
\tilde{R}_\beta=
V_\beta^\dagger
[I_2^{\otimes (2\nb)}\otimes Q]
V_\beta
\;.
\eeq

\begin{figure}[h]
    \begin{center}
    \includegraphics[scale=.60]{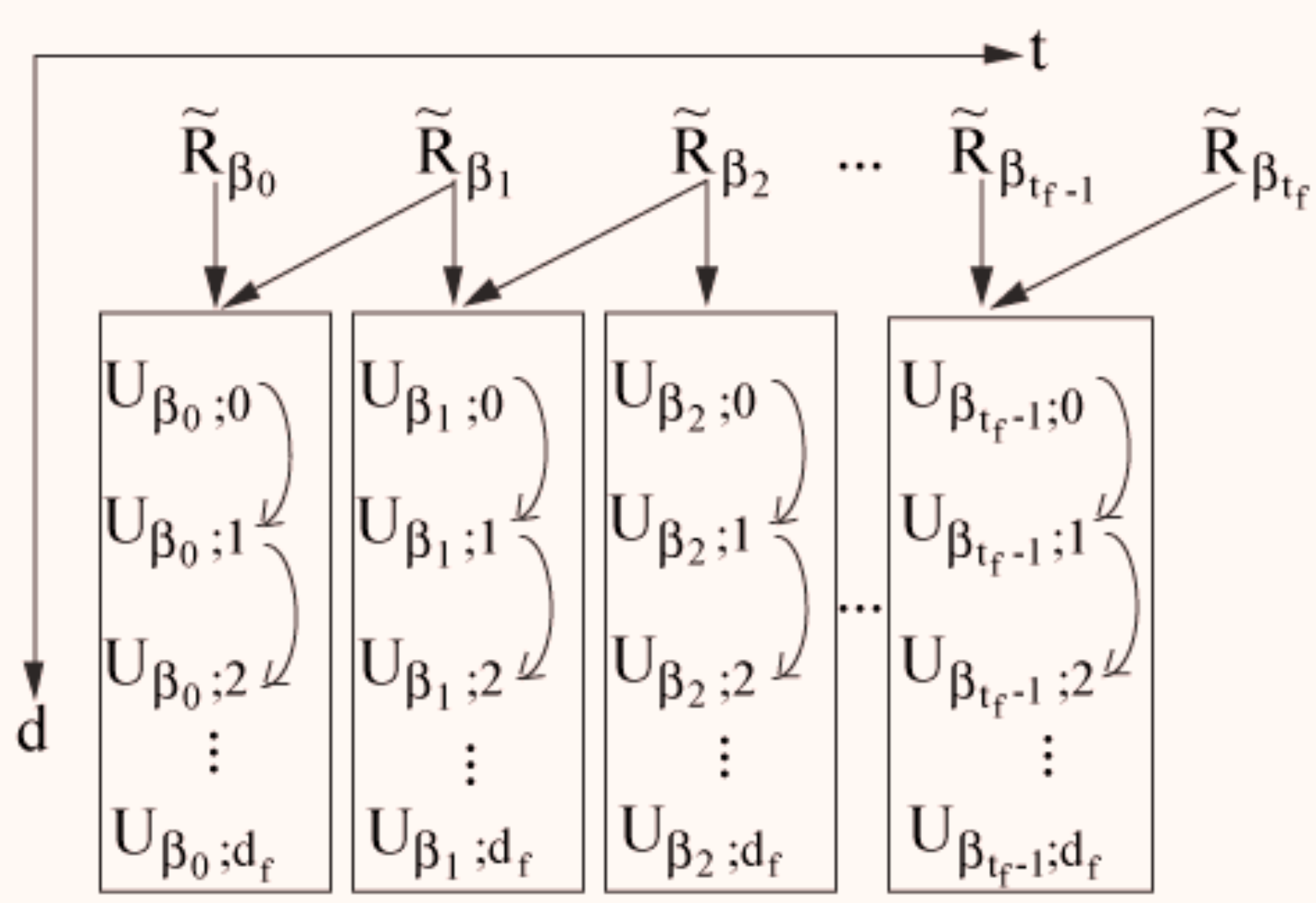}
    \caption{Pictorial representation
    of Grover recursion Eq.(\ref{eq-grover-recursion}).}
    \label{fig-grover-recursion}
    \end{center}
\end{figure}

Consider an annealing schedule
$\beta_0, \beta_1,
\ldots,\beta_{t_f}$.
For each ``time" $t\in\{0,1,\ldots,t_f\}$,
define a unitary matrix $U_{\beta_t;d_f}$
recursively as follows:
Let

\begin{subequations}
\label{eq-grover-recursion}
\beq
U_{\beta_t; 0} = I_2^{\otimes (2\nb + ac)}
\;,
\eeq
and

\beq
U_{\beta_{t}; d+1}=
U_{\beta_{t}; d}
\tilde{R}_{\beta_{t}}
U^\dagger_{\beta_{t}; d}
\tilde{R}_{\beta_{t+1}}
U_{\beta_{t}; d}
\;.
\eeq
\end{subequations}
where $d\in\{0,1,\ldots,d_f-1\}$.
See Fig.\ref{fig-grover-recursion}
for a pictorial representation
of this recursion.
We will call the parameter $d_f=1,2,\ldots$
which is the final level of
recursion,
the {\bf Grover depth}.

If $\calu$ is defined by

\beq
\calu=
U_{\beta_{t_f-1}; d_f}
\ldots
U_{\beta_{2}; d_f}
U_{\beta_{1}; d_f}
U_{\beta_{0}; d_f}
\;,
\eeq
and if we've done things right,
$\calu$ should satisfy

\beq
\calu\ket{\sqrt{\pi_{\beta_0}}}\otimes\ket{0}^\nb\otimes\ket{0}^{\otimes(ac)}
\approx
\ket{\sqrt{\pi_{\beta_{t_f}}}}\otimes\ket{0}^\nb\otimes\ket{0}^{\otimes(ac)}
\;.
\eeq

\end{document}